\documentclass[twocolumn,abs,prb,longbibliography]{revtex4-1}

\usepackage{graphicx}
\usepackage{dcolumn}
\usepackage{float}
\usepackage{amsmath}
\usepackage{bm}

\newcommand{\comment}[1]{}

\begin{document}


\title{Optical properties of a semi-Dirac material}

\author{J. P. Carbotte$^{1,2}$}
\email{carbotte@mcmaster.ca}
\author{K. R. Bryenton$^{3}$}
\author{E. J. Nicol$^{3}$}
\email{enicol@uoguelph.ca}
\affiliation{$^1$Department of Physics and Astronomy, McMaster
University, Hamilton, Ontario L8S 4M1, Canada}
\affiliation{$^2$The Canadian Institute for Advanced Research, Toronto, ON M5G 1Z8, Canada}
\affiliation{$^3$Department of Physics, University of Guelph,
Guelph, Ontario N1G 2W1, Canada} 
\date{\today}

\begin{abstract}{
Within a Kubo formalism, we calculate the absorptive part of the dynamic longitudinal conductivity $\sigma(\Omega)$ of a 2D semi-Dirac material. In the clean limit, we provide separate analytic formulas for intraband (Drude) and interband contributions for $\sigma(\Omega)$ in both the relativistic and nonrelativistic directions. At finite doping, in the relativistic direction, a sumrule holds between the increase in optical spectral weight in the Drude component and that lost in the interband optical transitions. For the nonrelativistic direction, no such sumrule applies. Results are also presented when an energy gap opens in the energy dispersion. Numerical results due to finite residual scattering are provided and analytic  results for the dc limit are derived. Energy dependence and possible anisotropy in the impurity scattering rate is considered. Throughout, we provide comparison of our results for $\sqrt{\sigma_{xx}\sigma_{yy}}$ with the corresponding results for graphene. A generalization of the 2D Hamiltonian to include powers of higher order than quadratic (nonrelativistic) and linear (relativistic) is considered. We also discuss the modifications introduced when an additional flat band is included via a semi-Dirac version of the $\alpha$-${\cal T}_3$ model, for which an $\alpha$ parameter tunes between the 2D semi-Dirac (graphene-like) limit and the semi-Dirac version of the dice or ${\cal T}_3$ lattice.
}
\end{abstract}

\maketitle

\section{Introduction}
The isolation of graphene\cite{Novoselov:2004,Novoselov:2005}, a two-dimensional, one-atom-thick carbon sheet with a honeycomb lattice, led to much experimental and theoretical work which rapidly uncovered many of its novel and exotic properties\cite{Geim:2007}. The idea and realization in the laboratory of 2D\cite{Kane:2005} and 3D\cite{Qi:2011,Hasan:2010} topological insulators soon followed. These are materials that are insulating in the bulk but have conducting  surface states. There now exists a vast literature on what has become to be known as Dirac materials. This includes the area of 3D Dirac and Weyl semimetals\cite{Vafek:2014} which are analogues of 2D graphene, and display such exotic phenomena as a chiral anomaly\cite{Hosur:2015,Ashby:2014,Hosur:2013,Gorbar:2014} and Fermi arcs on their surfaces\cite{Wan:2011,Potter:2014}.

Optical absorption experiments have provided a wealth of information on the dynamics of the charge carriers in a large class of materials including the high $T_c$ cuprate superconductors\cite{Basov:2005,Carbotte:2011}, graphene\cite{Li:2008,Carbotte:2010}, topological insulators\cite{Schafgans:2012}, as well as, Weyl and Dirac\cite{Chen:2015,Sushkov:2015,Xu:2016,Neubauer:2016,Chinotti:2016,Tabert:2016a,Tabert:2016b,Carbotte:2016} semimetals. In this paper, we consider the optical response of semi-Dirac 2D materials which display a linear-in-momentum (relativistic) dispersion along one direction and a quadratic one (non-relativistic) in the other direction. Such a dispersion was found in theoretical calculations of TiO$_2$/VO$_2$ nanostructures\cite{Pardo:2009} and in the semiconducting state of $\alpha$-(BEDT-TTF)$_{13}$ salts\cite{Katayama:2006}. It also arises in models of the merging of Dirac points\cite{Montambaux:2009a,Montambaux:2009b,Goerbig:2008} in two-dimensional crystals. Such merging has been observed in a Fermionic cold atom gas loaded into an optical lattice.\cite{Tarruell:2012}. Some properties of semi-Dirac semimetals have been elaborated on in many previous studies\cite{Banerjee:2012,Banerjee:2009}. Some examples include the effect of merging Dirac points on: the Floquet topological transition in graphene\cite{Delplace:2013}, the emergence of a Chern insulating state\cite{Huang:2015}, the dynamic polarization function and plasmons\cite{Pyatkovskiy:2016}, valley-selective Landau-Zener oscillations in p-n junctions\cite{Saha:2017}, and the magnetic susceptibility in an $\alpha$-${\cal T}_3$ model\cite{Piechon:2015}. Ziegler and Sinner\cite{Ziegler:2017} have calculated the interband ac conductivity $\sigma(\Omega)$ as a function of photon energy $\Omega$ in the clean limit. They find large anisotropies between $\sigma^{xx}_{\rm inter}(\Omega)$ and $\sigma^{yy}_{\rm inter}(\Omega)$. They also calculate a diffusion coefficient from the mean square displacement of the charge carriers and find that it is very different for relativistic than for the nonrelativistic dispersion curves.  In related work, Adroguer {\it et al.}\cite{Adroguer:2016} use both a Boltzmann formalism and a Kubo formula approach in a random impurity model to treat the optical intraband transitions. They find a resulting residual scattering rate which is not only dependent on energy but is, as well, highly anisotropic. The energy dependence is related to the variation with energy of the underlying carrier density of states. The anisotropy relates to the change in the dispersion curves from relativistic in the $y$-direction to nonrelativistic in the $x$-direction. Both effects are important but are also quite specific to the chosen impurity model. Large differences between 
$\sigma^{xx}(\Omega)$ and $\sigma^{yy}(\Omega)$
 imply that  for an electric field orientated in an arbitrary direction with respect to the $(x,y)$ axes, the conductivity will have a finite transverse, as well as, a longitudinal component.\cite{Sanderson:2018} This is in contrast to graphene which is isotropic. The anisotropy of the semi-Dirac model, which is not part of the pure Dirac case, has other consequences. For instance, the transmittance as a function of incident polarization angle\cite{Nualpijit:2018} 
is altered in an important way, as is the dichroism\cite{Oliva-Leyva:2016}. Another discussion of the effect of anisotropy on transport and other properties was given by Sriluckshmy {\it et al.}\cite{Sriluckshmy:2018}. They treat impurity scattering in a self-consistent Born approximation and emphasise that the resulting self-energy has a nonzero offdiagonal component which leads to an enlargement of the phase diagram in a semi-Dirac material to include a two-node Chern state which is not part of the clean case. In the present paper, we are mainly interested in the real (absorptive part) of the dynamic longitudinal conductivity
$\sigma^{xx}(\Omega)$ and $\sigma^{yy}(\Omega)$, first in the clean limit and it modification when weak impurity scattering is included.

In section II, we discuss the continuum Hamiltonian which we employ to calculate the conductivity in a Kubo formalism. In section III, we consider the clean limit and provide simple analytic formulas for the interband contribution to the ac optical conductivity. In semi-Dirac
$\sigma^{xx}_{\rm inter}(\Omega)$ goes as the square root of the photon energy $\Omega$ and $\sigma^{yy}_{\rm inter}(\Omega)$ has the inverse dependence. Both depend on the material parameters of mass $m$ and velocity $v$ defining the Hamiltonian. Our formulas agree with results presented in the very recent paper of Nualpijit {\it et al.}\cite{Nualpijit:2018} . As noted in that paper, the above material parameters cancel out of the square root of the product
$\sigma^{xx}_{\rm inter}(\Omega)\sigma^{yy}_{\rm inter}(\Omega)$
and consequently this quantity is universal and constant as in graphene, independent of photon energy.
Analytic expressions are also presented for the intraband contribution to the optical conductivity (the Drude contribution). We find that 
$\sigma^{xx}_{\rm intra}(\Omega)\propto \mu^{3/2}$, 
with $\mu$ the chemical potential, while $\sigma^{yy}_{\rm intra}(\Omega)\propto \mu^{1/2}$. Here again, the material parameters drop out of the square root of the product
$\sigma^{xx}_{\rm intra}(\Omega)\sigma^{yy}_{\rm intra}(\Omega)$ 
and this quantity is linear in chemical potential as in graphene. In graphene, there is a the famous conservation law\cite{Frenzel:2014} which states that the increase in optical spectral weight in the Drude contribution, $W_{\rm intra}$, due to increased doping (larger $\mu$) agrees exactly with the loss in optical spectral weight seen in the interband conductivity $W_{\rm inter}$. This has its origin in the increase in Pauli blocking of interband transitions as $\mu$ is increased. Remarkably, we find that a sum rule still applies to $\sigma^{yy}$ but not to $\sigma^{xx}$.
In the case of semi-Dirac, $W_{\rm inter}$ is smaller than 
$W_{\rm intra}$ 
by a factor of $2/3$ for $W_{\rm intra}=\sqrt{W^{xx}_{\rm intra} W^{yy}_{\rm intra} }$ and $W_{\rm inter}=\sqrt{W^{xx}_{\rm inter} W^{yy}_{\rm inter} }$.
We also study the temperature variation of the Drude optical spectral weight at charge neutrality as well as for finite doping.

In section IV, we consider the case when a gap is included in the model Hamiltonian, namely adding a term $\sigma_3\Delta$, with $\sigma_3$ the usual third Pauli matrix. Analytic formulas are obtained which reduce properly to the known $\Delta=0$ results. 
$\sigma^{xx}_{\rm inter}(\Omega)$ 
is found to go as $[\Omega^2-4\Delta^2]^{1/4}$ 
above the gap edge $\Omega=2\Delta$  while
$\sigma^{yy}_{\rm inter}(\Omega)$ 
behaves as the inverse. The square root of the product agrees with known results for graphene when the appropriate limit is considered. An additional study of the conductivity arising from the intraband transitions is also presented. The dependence on chemical potential $\mu$ is found to be $(\mu^2-\Delta^2)^{5/4}/\mu$ and $(\mu^2-\Delta^2)^{3/4}/\mu$ for $xx$ and $yy$ directions, respectively.

In section V, we go beyond the clean limit. To get a first understanding of the generic modifications that arise, it is sufficient to include residual scattering in the simplest possible approximation, namely, a constant scattering rate $\Gamma$. When this is introduced in the one loop approximation Kubo formula used in all our work, we compute the real part of the dynamic conductivity. We find that in this model, a simple scaling applies and 
$\sigma^{xx}(T,\mu,\Omega) \equiv (e^2/h)\sqrt{\Gamma/mv^2}\bar\sigma^{xx}(\bar T,\bar\mu,\bar\Omega) $ and
$\sigma^{yy}(T,\mu,\Omega) \equiv (e^2/h)\sqrt{mv^2/
\Gamma}\bar\sigma^{yy}(\bar T,\bar\mu,\bar\Omega) $, where $\bar T=T/\Gamma$, $\bar \mu=\mu/\Gamma$, and
$\bar \Omega=\Omega/\Gamma$, which implies that at zero temperature $T=0$, we have a single family of curves labelled by $\bar\mu$ for $\bar\sigma(0,\bar\mu,\bar\Omega)$ versus $\Omega/\Gamma$. The prefactors in front of $\bar\sigma^{xx}$
and $\bar\sigma^{yy}$ are different and in particular in the dc limit
$\sigma^{xx}_{dc}\sim\sqrt{\Gamma/mv^2}$
while
$\sigma^{yy}_{dc}$
has the inverse variation. These dependencies cancel out of the square root of the product $\sigma^{xx}(0,\mu,\Omega)\sigma^{yy}(0,\mu,\Omega)$
which becomes an universal  constant independent of any material parameter, including the residual scattering rate $\Gamma$, as found in graphene.

In section VI, we consider a generalization of the semi-Dirac model Hamiltonian in which the exponents on $k_x$ and $k_y$ are taken to be a general positive integer $n$ and $s$, respectively. The case, $n=s=1$ corresponds to graphene and $n=2$ and $s=1$ gives semi-Dirac. We show that for this model in the clean limit $\sigma^{xx}_{\rm inter}\propto \Omega^{-{1\over n}+{1\over s}}$
while 
$\sigma^{yy}_{\rm inter}\propto \Omega^{-{1\over s}+{1\over n}}$.
For any $n=s$, the dependence on photon energy drops out of these quantities as it does in the square root of the product. This quantity is constant and universal independent of material parameters, but has a different magnitude as $n=s$ is varied.

In section VII, we provide a brief description of the generalization of our work to include a flat band in a semi-Dirac $\alpha$-${\cal T}_3$ model.\cite{Piechon:2015} The role of flat bands has been of considerable recent interest and evidence for $\alpha$-${\cal T}_3$ behavior has been seen in experiments on Hg$_{1-x}$Cd$_x$Te which exhibits a nearly flat band and cones at a critical doping of $x\simeq 0.17$.\cite{Illes:2015,Malcolm:2015,Illes:2016,Malcolm:2016,Orlita:2014}

In section VIII, we give a brief summary of our main results and draw conclusions.

\section{Theoretical Background}
\begin{figure}
\includegraphics[width=0.9\linewidth]{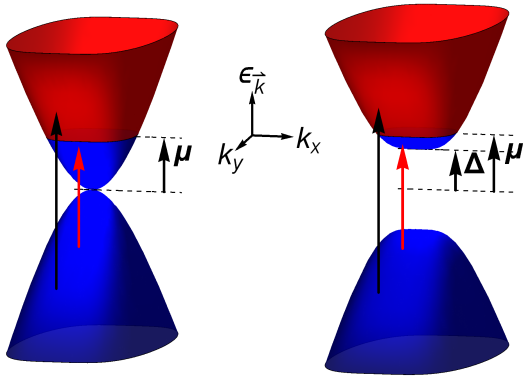}
\caption{The energy dispersion in the semi-Dirac model (left) and the same with an energy gap $\Delta$ measured from charge neutrality (right). Pictures are shown for finite doping $\mu$ with blue representing occupied states and red, unoccupied states. Red arrows show examples of optical transitions that are blocked by Pauli exclusion and black arrows represent transitions that result in optical absorption.
}\label{fig1}
\end{figure}

In this work, we consider a low energy Hamiltonian which embodies a semi-Dirac
behavior about a particular Dirac point. For graphene, the Hamiltonian about a Dirac point is the Weyl Hamiltonian which is a $2\times 2$ matrix with zeroes on the diagonal and terms linear in $k=|{\bm k}|$, specifically $\hbar v (k_x\pm i k_y)=\hbar v ke^{\pm i\phi}$, on the offdiagonal elements. The resulting linear dispersion $\epsilon_k=\pm\hbar v k$ is referred to as Dirac-like behavior or ``relativistic''. Here, $v$ is the Fermi velocity. In the semi-Dirac case, the Hamiltonian is modified to be
\begin{equation}
\hat H=\left(\begin{array}{cc}
0 & \displaystyle\frac{\hbar^2k_x^2}{2m}-i\hbar vk_y\\
\displaystyle\frac{\hbar^2k_x^2}{2m}+i\hbar vk_y & 0
\end{array}\right) ,
\label{eq:Hmatrix}
\end{equation}
where the $x$-direction now has a quadratic-in-$k$ behaviour with a mass $m$ as one finds for free electrons (non-relativistic). Thus, the semi-Dirac nature is that one direction ($y$ in this case) is linear in $k$ and the other direction is quadratic in $k$. This gives a dispersion relation of 
$\epsilon_k=\pm\sqrt{(\hbar^2k_x^2/2m)^2+\hbar^2v^2k^2_y}$. 
This dispersion is plotted in Fig.~\ref{fig1} on the left. We use the Hamiltonian of Eq.~(\ref{eq:Hmatrix}) to evaluate the consequences of such a combination on the optical or dynamical conductivity as a function of incident photon frequency $\Omega$.

Using $\hat G({\bm k},z)= [z\hat I-\hat H]^{-1}$ to evaluate the matrix Green function from the Hamiltonian 
for complex variable $z$ ($\hat I$ is the identity matrix), the spectral function matrix $\hat A({\bm k},\omega)$ can be evaluated from $\hat A({\bm k},\omega)=-2 {\rm Im} \hat G({\bm k}, z\to \omega+i0^+)$. This then enters the Kubo formula for the real part of the conductivity (the absorptive part), written at finite temperature $T$ as:
\begin{eqnarray}
\sigma^{ij}(T,\Omega)=& \displaystyle N_f\hbar\frac{ e^2}{2\Omega}\int^{+\infty}_{-\infty}{d\omega\over 2\pi} [f(\omega) - f(\omega+\Omega)]\nonumber\\
&\displaystyle\times\int \frac{d^2{k}}{4\pi^2} {\rm Tr}[\hat v_i \hat A({\bm k},\omega) \hat v_j \hat A({\bm k},\omega+\Omega)]
\label{eq:condformula}
\end{eqnarray}
where the velocity matrix is given by $\hbar\hat v_i =\partial \hat H/\partial k_i$, with $i$ (and $j$) referring to $x$ or $y$. Here, $f(\omega)$ is the Fermi-Dirac function $f(\omega)=1/(1+{\rm exp}[(\omega-\mu)/k_BT])$, where $\mu$ is the chemical potential measured from the charge neutrality point and taken to be positive (see Fig.~\ref{fig1}), and $k_B$ is the Boltzmann constant. $N_f$ is a degeneracy factor which would account for spin degeneracy $g_s$ and valley degeneracy or multiple Weyl/Dirac points $g_v$. In this work, we will take $N_f=g_sg_v=1$ ({\it i.e.,} the conductivity will be quoted per spin and per valley). However, in comparisons with graphene, it must be remembered that, in general, the results for graphene include the degeneracy factor of $N_f=4$. While we focus on the absorptive part of the conductivity in this work, the imaginary part of the conductivity is related to this through a Kramers-Kronig 
 transformation.\cite{Tabert:2012}

\section{Clean limit}
We start by considering the clean limit at $T=0$ which captures the essence of the behavior of the conductivity. In the absence of impurity scattering and other possible many body effects, the spectral function matrix evaluated from the Hamiltonian in Eq.~(\ref{eq:Hmatrix})  will be composed of quantities involving delta functions and as a result analytical results can be obtained in a straightforward manner. Examples of such type of calculations following this approach may be found in Refs.~[\onlinecite{Nicol:2008,Tabert:2012,Stille:2012}] and so we only provide the final results here. For zero temperature, the interband contribution, which provides the background conductivity at finite photon frequency $\Omega$, is given for the two longitudinal components ($xx$ and $yy$) as:
\begin{eqnarray}
\sigma_{\rm inter}^{xx}(\Omega) &= \displaystyle{e^2\over h}\sqrt{{\Omega\over mv^2}} {1\over 5G}\theta(\Omega-2\mu),\label{eq:intercleanxx}\\
\sigma_{\rm inter}^{yy}(\Omega) &= \displaystyle{e^2\over h}\sqrt{{mv^2\over\Omega}} {\pi G\over 6}\theta(\Omega-2\mu),
\label{eq:intercleanyy}
\end{eqnarray}
where $G\approx 0.8346$ is the Gauss constant (see the Appendix for further information).
The $\theta(\Omega-2\mu)$ is the Heaviside function where $\theta(x>0)=1$ and $\theta(x<0)=0$. It results from the Pauli exclusion principle applied through the combination of Fermi-Dirac functions in Eq.~(\ref{eq:condformula}) where only vertical transitions from occupied to unoccupied states are allowed as the photon is essentially a ${\bm q}=0$ probe, transferring no momentum. A typical transition is illustrated by the long black arrow in Fig.~\ref{fig1}, with the arrow shown in red, disallowed. Hence, the lowest interband transition, at $T=0$, results in absorption starting at $\Omega=2\mu$.
From Eqs.~(\ref{eq:intercleanxx}) and (\ref{eq:intercleanyy}), we note that $\sigma^{xx}(\Omega)$ goes like $\sqrt{\Omega}$ and 
$\sigma^{yy}(\Omega)$ goes like $1/\sqrt{\Omega}$. If we take the square root of the product these two conductivities, the dependence on photon energy drops out entirely and 
$\sigma_{\rm inter}(\Omega)\equiv\sqrt{\sigma^{xx}_{\rm inter}\sigma^{yy}_{\rm inter}}$ is a constant:
\begin{equation}\displaystyle
\sigma_{\rm inter}(\Omega)={e^2\over h}\sqrt{{\pi\over 30}}\theta(\Omega-2\mu).
\label{eq:sigmainter}
\end{equation}
Note, as mentioned before, we have not included a degeneracy factor $N_f$ which for graphene is 4 (accounting for spin and valley degeneracy). If we take this fact into consideration when comparing graphene and semi-Dirac, we see that for the semi-Dirac case, there also exists a universal interband background as in graphene but this applies only to the square root of the product $\sigma_{\rm inter}(\Omega)=\sqrt{\sigma^{xx}_{\rm inter}\sigma^{yy}_{\rm inter}}$  and its value is 17.6\% smaller than for graphene (single valley and single spin) which corresponds to replacing the square root in Eq.~(\ref{eq:sigmainter}) by $\pi/2$.
Ziegler and Sinner\cite{Ziegler:2017} have found the same low energy behavior of the interband conductivities through numerical tight binding calculations performed over the Brillouin zone, but in their case, they have deviations at high frequency due to the presence of a van Hove singularity which is not included in our low energy continuum model. Consequently, the universal background would change from the continuum model approximation.

\begin{figure}
\includegraphics[width=0.9\linewidth]{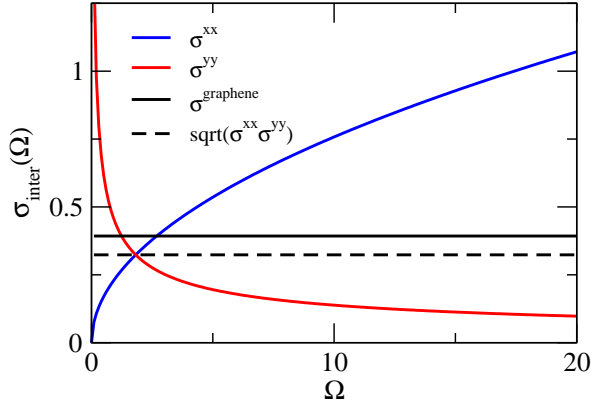}
\caption{The real part of the interband longitudinal conductivity
as a function of the photon energy $\Omega$ for the $xx$ and $yy$ cases. Comparison is made with graphene per single valley and single spin, {\it i.e.}, $\sigma^{\rm graphene}=\pi e^2/(8h)=0.393e^2/h$. $\sigma_{\rm inter}(\Omega)=\sqrt{\sigma^{xx}_{\rm inter}\sigma^{yy}_{\rm inter}} = (e^2/h)\sqrt{\pi/30}=0.324 e^2/h$ is also plotted for comparison. The conductivity is normalized by $e^2/h$ and the photon energy for the $xx$ and $yy$ cases is normalized by $mv^2$.
}\label{fig2}
\end{figure}

In Fig.~\ref{fig2}, we compare results for results for $\sigma_{\rm inter}^{xx}$ and for $\sigma_{\rm inter}^{yy}$ with the case of graphene ($N_f$ has been taken to be 1 in all cases). We see that $\sigma^{xx}_{\rm inter}$  and $\sigma^{yy}_{\rm inter}$ are not universal like graphene due to the product of the material parameters $mv^2$ which scales the curves. Likewise, these two conductivities are not constant as a function of frequency. However, $\sigma_{\rm inter}(\Omega)=\sqrt{\sigma^{xx}_{\rm inter}\sigma^{yy}_{\rm inter}}$ is universal and displays a constant interband background.

We next consider the intraband contribution to the conductivity in the clean limit, which we find to be:
\begin{eqnarray}
\sigma_{\rm intra}^{xx}(\Omega) &= \displaystyle{e^2\over h}
{12\over 5G}{\mu^{3/2}\over\sqrt{2mv^2}} \delta(\Omega),\label{eq:intrabandcleana}
\\
\sigma_{\rm intra}^{yy}(\Omega) &= \displaystyle{e^2\over h}
{2\pi G\over 3}\sqrt{2mv^2\mu} \delta(\Omega).
\label{eq:intrabandcleanb}
\end{eqnarray}
Again, the material parameters of $m$ and $v$ drop out of the square root of the product of these two quantities and we get
\begin{eqnarray}
\sigma_{\rm intra}(\Omega)&\equiv&\sqrt{\sigma^{xx}_{\rm intra}\sigma^{yy}_{\rm intra}}\\
                        &=&\displaystyle {e^2\over h} \sqrt{{8\pi\over 5}}\mu 
\delta(\Omega),
\label{eq:sqrtintra}
\end{eqnarray}
which is linear in the chemical potential $\mu$ as in graphene, while individually $\sigma_{\rm intra}^{xx}$ goes like $\mu^{1/2}$ and $\sigma_{\rm intra}^{yy}$ like 
$\mu^{3/2}$. The squareroot of the product shown in Eq.~(\ref{eq:sqrtintra}) varies exactly as in graphene except for a modification in magnitude by a factor of $(2/\pi)\sqrt{8\pi/5}\simeq 1.43$. Thus, there is more optical spectral weight in the Drude (or intraband) component for semi-Dirac than for graphene for the same value of chemical potential $\mu$.

The intraband optical spectral weight is defined as the area under the intraband conductivity curve: $W_{\rm intra}^{ii}=\int_0^\infty \sigma_{\rm intra}^{ii}(\Omega)d\Omega$. The amount of optical spectral weight that resides in the intraband transitions (Drude contribution) is 
\begin{equation}
W^{xx}_{\rm intra}=\displaystyle {e^2\over h}{6\over 5G}{\mu^{3/2}\over\sqrt{2mv^2}}
\end{equation}
and 
\begin{equation}
W^{yy}_{\rm intra}=\displaystyle {e^2\over h}{\pi G\over 3}{\sqrt{2mv^2\mu}}.
\end{equation}
For graphene, a famous sum rule applies to the transfer of optical spectral weight from the interband background to the Drude as the chemical potential $\mu$ is increased due to doping away from charge neutrality. The {\it missing} interband optical spectral weight that is transferred from the interband component is defined by 
$W^{ii}_{\rm inter}=\int_0^\infty[\sigma_{\rm inter}^{ii}(\Omega,\mu=0)-\sigma_{\rm inter}^{ii}(\Omega,\mu)]d\Omega=\int_0^{2\mu} \sigma_{\rm inter}^{ii}(\Omega,\mu=0)d\Omega$, where the latter integral is for $T=0$. In graphene
$W_{\rm inter}$ and $W_{\rm intra}$ are equal. Here, we find that this sum rule remains valid in semi-Dirac for the $yy$ conductivity  but is violated for the $xx$.
The missing optical spectral weight in $\sigma^{xx}_{\rm inter}$ and $\sigma^{yy}_{\rm inter}$ due to a finite chemical potential $\mu$ is 
\begin{equation}
W^{xx}_{\rm inter}=\displaystyle {e^2\over h}{8\over 15 G}
{\mu^{3/2}\over\sqrt{2mv^2}},
\end{equation}
and 
\begin{equation}
W^{yy}_{\rm inter}=\displaystyle {e^2\over h}{\pi G\over 3}{\sqrt{2 mv^2\mu}}.
\end{equation}
This missing spectral weight is a result of Pauli blocking. The finite occupation of the conduction band up to energy $\mu$ means that  optical transitions from valence band to conduction band with photon energy less than $2\mu$ are no longer possible. While direction-specific inter and intra spectral weights shown above have the same dependence on $\mu$, specifically $\mu^{3/2}$ for $xx$ and $\mu^{1/2}$ for $yy$, their magnitude need not be the same. We find
\begin{equation}
\displaystyle {W^{xx}_{\rm inter}\over W^{xx}_{\rm intra}}= {4\over 9}
\end{equation}
and 
\begin{equation}
\displaystyle {W^{yy}_{\rm inter}\over W^{yy}_{\rm intra}}=1,
\end{equation}
so that a sumrule does indeed apply to the $yy$ direction but not to the $xx$ where more spectral weight is found in the Drude than is lost in the interband background by a factor of 9/4, which is more than a factor of two.
Finally, defining $W_{\rm intra}\equiv\sqrt{W^{xx}_{\rm intra}W^{yy}_{\rm intra}}$ and $W_{\rm inter}\equiv\sqrt{W^{xx}_{\rm inter}W^{yy}_{\rm inter}}$, then 
\begin{equation}
\displaystyle {W_{\rm inter}\over W_{\rm intra}}={2\over 3}.
\end{equation}

\begin{figure}
\includegraphics[width=0.9\linewidth]{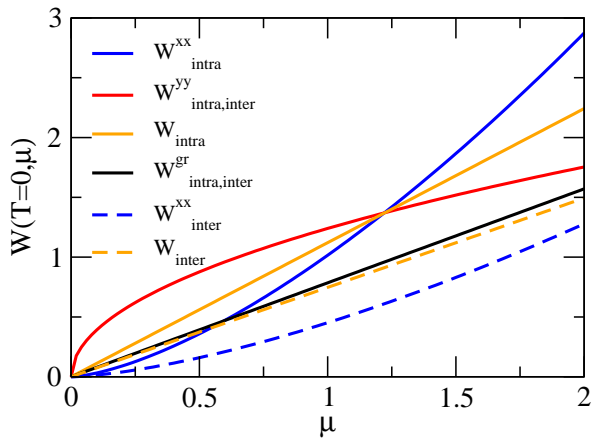}
\caption{Optical spectral weight as a function of the chemical potential. The chemical potential $\mu$ and the spectral weight are normalized by $mv^2$ and $e^2mv^2/h$, respectively. Shown are the intraband and missing interband spectral weights as discussed in the text.
}\label{fig3}
\end{figure}

To emphasize these results, we plot in Fig.~\ref{fig3}  the optical spectral weight that resides in the Drude $W^{yy}_{\rm intra}$ (red curve) at zero temperature as a function of chemical potential $\mu$ and compare with the $xx$ direction
$W^{xx}_{\rm intra}$ (solid blue curve). In units of $e^2mv^2/h$, $xx$ goes like $1.02(\mu/mv^2)^{3/2}$ while $yy$ goes as $1.24\sqrt{\mu/mv^2}$. The square root of the product of these two quantities $W_{\rm intra}$ 
is not dependent on the material parameters $m$ and $v$. It is linear in $\mu$ and equal to $1.12\mu$ in units of $e^2/h$ (solid orange line). The linear in $\mu$ dependence of $W_{\rm intra}$ in semi-Dirac is the same dependence as found in graphene $W^{\rm gr}_{\rm intra}/(e^2/h)=\pi\mu/4$ (black line). The slope of these two variations are, however, quite different, with the semi-Dirac case larger by a factor of 1.43.

Next we compare with the missing optical spectral weight in the interband background that results from doping, {\it i.e.}, a finite $\mu$ value. 
$W^{yy}_{\rm inter}$ is identical to 
$W^{yy}_{\rm intra}$ (red curve) and in this direction, the missing spectral weight in the background is simply transferred to the Drude, as is found in graphene. However, while 
$W^{xx}_{\rm inter}$ (dashed blue curve) has the same $\mu^{3/2}$ dependence on chemical potential as does its intraband counterpart, it is much smaller and equal to $0.452\mu^{3/2}$ rather than $1.01\mu^{3/2}$, in the units of Fig.~\ref{fig3}. In this direction, the amount of spectral weight lost in the interband background is less than half the amount gained in the intraband (Drude) part. Consequently, no conservation of optical spectral weight holds. This will also be the case for the composite quantities 
  $W_{\rm intra}$ (solid orange)
and $W_{\rm inter}$ 
(dashed orange).
Both vary linearly with $\mu$ but they have a very different slope of 1.12 and 0.747, respectively.

Remaining in the clean limit, we now turn to a discussion of the temperature and doping dependence ($\mu$-dependence) of the optical spectral weight residing under the Drude peak. This comes entirely from the intraband optical transitions. In Eqs.~(\ref{eq:intrabandcleana})-(\ref{eq:sqrtintra}), the limit of zero temperature was taken. If we instead stay at finite temperature, we get for the Drude spectral weight
\begin{equation}
W^{yy}_{\rm intra}=\displaystyle {e^2\over h}{\pi G\over 3}\sqrt{2mv^2}
\int_{-\infty}^{\infty}\biggl(-{\partial f(\omega)\over\partial\omega}\biggr)\sqrt{|\omega|}d\omega,
\end{equation}
which can be rewritten as ($k_B=1$)
\begin{equation}
W^{yy}_{\rm intra}=\displaystyle {e^2\over h}{\pi G\over 3}\sqrt{mv^2 T}
\int_{0}^{\infty}\sqrt{x}I(x,\mu,T)dx,
\label{eq:wyyintra}
\end{equation}
with
\begin{equation}
I(x,\mu,T)=\displaystyle {1\over\cosh^2(x-{\mu\over 2T})}+{1\over\cosh^2(x+{\mu\over 2T})}.
\end{equation}
Similarly,
\begin{equation}
W^{xx}_{\rm intra}=\displaystyle {e^2\over h}{6\over 5G}{1\over\sqrt{mv^2}}T^{3/2}
\int_{0}^{\infty}x^{3/2}I(x,\mu,T)dx.
\label{eq:wxxintra}
\end{equation}
In graphene, we can get an analytic result for the optical spectral weight in the Drude (intraband optical transitions) at finite temperature and chemical potential:\cite{Gusynin:2009}
\begin{eqnarray}
W^{\rm gr}_{\rm intra}&=\displaystyle {e^2\over h}{\pi T\over 4}
\int_{0}^{\infty}xI(x,\mu,T)dx,\\
&=\displaystyle{e^2\over h}{\pi T\over 2}\ln\biggl[2\cosh\biggl({\mu\over 2T}\biggr)\biggr].
\label{eq:wgintra}
\end{eqnarray}
While Eqs.~(\ref{eq:wyyintra}) and (\ref{eq:wxxintra}) are valid for any doping, they simplify at charge neutrality $\mu=0$ in which limit we obtain:
\begin{eqnarray}
W^{xx}_{\rm intra}&=\displaystyle 1.44 {e^2\over h}{6\over 5G}{1\over\sqrt{mv^2}} T^{3/2},\\
W^{yy}_{\rm intra}&=\displaystyle 1.52 {e^2\over h}{\pi G\over 3}\sqrt{mv^2T},
\end{eqnarray}
and this gives
\begin{equation}
W_{\rm intra}=\displaystyle 1.65{e^2\over h}T.
\end{equation}
This last quantity is linear in temperature as for graphene for which $W^{\rm gr}_{\rm intra}=1.09(e^2/h)T$ with the valley and spin degeneracy factor set equal to one for direct comparison with the semi-Dirac case. The semi-Dirac case is larger in magnitude by about 50\%.

\begin{figure}
\includegraphics[width=0.9\linewidth]{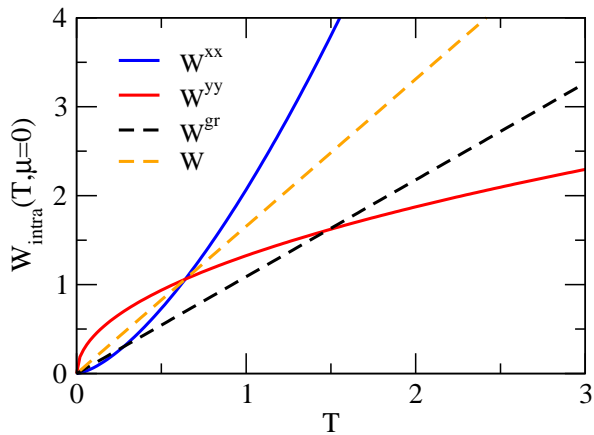}
\caption{The Drude weight or intraband optical spectral weight for $\mu=0$ versus temperature $T$. With $T$ normalized by $mv^2$, the Drude weight is normalized by $e^2mv^2/h$.
}\label{fig4}
\end{figure}

In Fig.~\ref{fig4}, we plot these spectral weight variations with temperature at
 charge neutrality. $W^{yy}_{\rm intra}$, in units of $e^2mv^2/h$, 
goes like the square root of temperature with proportionality constant 1.33 (solid red curve). $W^{xx}_{\rm intra}$ goes instead like $T^{3/2}$ with proportionality constant 2.07 (solid blue curve). The square root of the product of these two quantities has units of $e^2/h$ and is linear in $T$ with slope 1.65 (dashed orange) to be compared with graphene which is also linear (dashed black) but with a much smaller slope.

\begin{figure}
\includegraphics[width=0.9\linewidth]{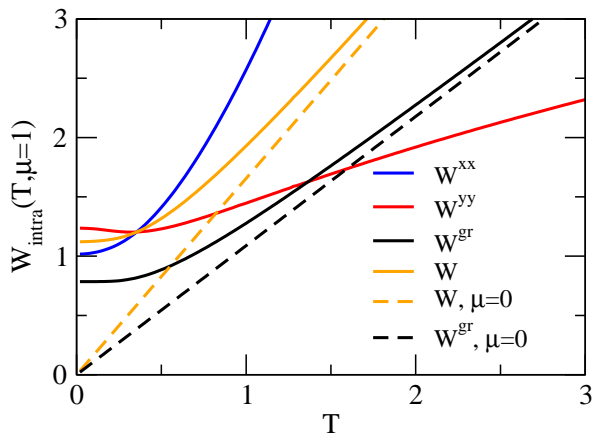}
\caption{The Drude or intraband optical spectral weight for finite $\mu=1$ as a function of temperature $T$. Units are as in Fig.~\ref{fig4}.
}\label{fig5}
\end{figure}

In Fig.~\ref{fig5}, we return to the case of Eqs.~(\ref{eq:wyyintra})-(\ref{eq:wgintra}) where both $T$ and $\mu$ are finite. We present results for a finite chemical potential value $\mu=1$, as well as two curves for $\mu=0$ for comparison with the finite $\mu$ results. The Drude spectral weight in the $yy$ direction of our semi-Dirac model for $\mu=1$ is given by the solid red curve which starts at a finite value at $T=0$ and has a very shallow minimum before rising slowly with increasing temperature $T$ to merge with the  $\mu=0$ solid red curve shown in  Fig.~\ref{fig4} which goes like $\sqrt{T}$. The solid blue curve is for $W^{xx}_{\rm intra}$ for $\mu=1$. It also starts at a finite value for $T=0$ and rises with increasing $T$ eventually to match up with the corresponding $\mu=0$ of Fig.~\ref{fig4} which goes as $T^{3/2}$. The square root of the product of these two quantities is shown as the solid orange curve which starts at a finite value at $T=0$. It falls between the solid red and the solid blue curves and asymptotically at large temperature is linear in $T$. The other curves are for comparison. The solid black curve is for graphene with $\mu=1$ while the dashed black curve is for $\mu=0$ (reproduced from Fig.~\ref{fig4}). The $\mu=0$ curve starts from zero, of course, and is linear in $T$ and the solid black merges with the $\mu=0$ case as $T$ increases. The dashed orange curve (also shown in Fig.~\ref{fig4}) is for $W_{\rm intra}=\sqrt{W^{yy}_{\rm intra}W^{xx}_{\rm intra}}$ in semi-Dirac with $\mu=0$ and is linear in $T$. While the magnitudes of these variations is changed as compared with graphene, the relationship between $W_{\rm intra}$ at $\mu=0$ and $\mu=1$ is very much the same as is found in graphene.

The optical spectral weight in the intraband transitions can be determined in experiments. Frenzel {\it et al.}\cite{Frenzel:2014} have measured this quantity in graphene at several values of carrier density created through a change in gate voltage. An optical pump probe technique in the terahertz is employed to create high energy carriers and to measure their absorption as they relax to equilibrium. In this way, these authors confirmed the temperature and carrier dependence of the Drude spectral weight expected in theory.\cite{Muller:2009,Gusynin:2009} Other relevant experiments exist in the pyrochlore Eu$_2$Ir$_2$O$_7$ based on optical data.\cite{Sushkov:2015} They find that the temperature dependence of the free carrier response is that expected for a Weyl semimetal. Note that for experiments at finite doping, the chemical potential will have a dependence on temperature  because the electronic density of states is not constant, and this needs to be included.
 
\begin{figure}
\includegraphics[width=0.9\linewidth]{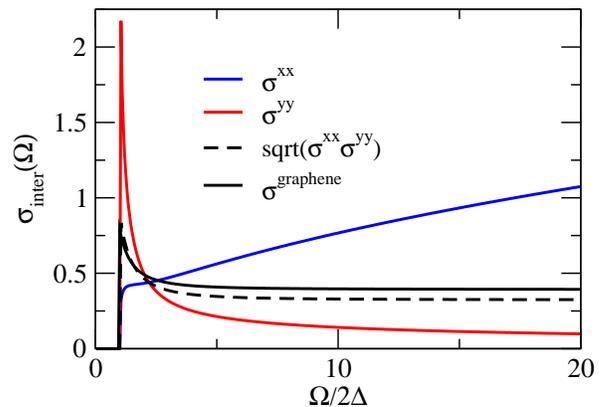}
\caption{The real part of the interband conductivity
as a function of the photon energy $\Omega/2\Delta$. Here, the presence of the energy gap $\Delta$ causes an abrupt onset at $2\Delta$ for $\mu=0$. These curves are plotted such that $\sigma^{xx}$ and $\sigma^{yy}$
are normalized by $(e^2/h)\sqrt{2\Delta/mv^2}$ and $(e^2/h)\sqrt{mv^2/2\Delta}$, respectively. Also shown is the
$\sigma^\Delta=\sqrt{\sigma^{xx}\sigma^{yy}}$ which is normalized by $e^2/h$. The solid black curve is the case for graphene with $N_f=1$ for comparison, again normalized by $e^2/h$.
}\label{fig6}
\end{figure}

\section{Including an energy gap}

In the case of graphene an onsite potential energy difference between the $A$ and $B$ sublattices of the graphene honeycomb lattice structure enters the Hamiltonian as a term $\Delta\sigma_z$, with $\sigma_z$ the two-dimensional Pauli matrix of 1 and -1 on the diagonal. $\Delta$ is an energy which gives rise to a gap in the dispersion $\epsilon_k=\pm\sqrt{(\hbar v k)^2+\Delta^2}$ and hence is referred to as a gap parameter here which has been widely used in the literature.\cite{Hill:2011,Hill:2013,Gusynin:2005,Gusynin:2009}  The equivalent statement for the semi-Dirac case modifies the Hamiltonian of Eq.~(\ref{eq:Hmatrix}) to be:
\begin{equation}
\hat H=\left(\begin{array}{cc}
\Delta & \displaystyle\frac{\hbar^2k_x^2}{2m}-i\hbar vk_y\\
\displaystyle\frac{\hbar^2k_x^2}{2m}+i\hbar vk_y & -\Delta
\end{array}\right) ,
\label{eq:Hmatrixgap}
\end{equation}
and the energy dispersion is modified to $E\equiv E_k=\pm\sqrt{\epsilon^2+\Delta^2}$,
where $\epsilon\equiv\epsilon_k=\pm\sqrt{(\hbar^2k_x^2/2m)^2+\hbar^2v^2k^2_y}$. We will favour the notation without the subscript $k$ for simplicity. This dispersion is plotted on the right in Fig.~\ref{fig1}.

Performing the calculations for this case, we arrive at the final analytical  forms:
\begin{eqnarray}
\sigma_{\rm inter}^{xx}(\Omega) =& \displaystyle{e^2\over h}{1\over 5G} 
\biggl[{\Omega^2-(2\Delta)^2\over(mv^2)^2}\biggr]^{1/4}
\biggl[1+6\biggl({\Delta\over\Omega}\biggr)^2\biggr]
\nonumber\\
&\times\theta(\Omega-2\Delta)\theta(\Omega-2\mu),\\
\sigma_{\rm inter}^{yy}(\Omega) =& \displaystyle{e^2\over h}{\pi G\over 6} 
\biggl[{(mv^2)^2\over \Omega^2-(2\Delta)^2}\biggr]^{1/4}
\biggl[1+8\biggl({\Delta\over\Omega}\biggr)^2\biggr]
\nonumber\\
&\times\theta(\Omega-2\Delta)\theta(\Omega-2\mu).
\label{eq:interbandgap}
\end{eqnarray}
The extra $\theta(\Omega-2\Delta)$ represents that the lack of states within the gap in the band structure results in no absorption below $2\Delta$ and for $\mu>\Delta$, there will be no absorption up to $2\mu$ due to the additional Pauli blocking (see the arrows in Fig.~\ref{fig1}, right side).
Both of these forms reduce properly to Eqs.~(\ref{eq:intercleanxx})
and (\ref{eq:intercleanyy}) in the limit of $\Delta\to 0$, as expected.
Note that in the $xx$ and $yy$ directions, the conductivity at the edge $\Omega=2\Delta$ goes like $[\Omega^2-(2\Delta)^2]^{1/4}$ and $[\Omega^2-(2\Delta)^2]^{-1/4}$, respectively. In Fig.~\ref{fig6}, we show the form of these interband conductivities with a finite gap. The square root of the product of the conductivities $\sigma^\Delta_{\rm inter}=\sqrt{\sigma^{xx}_{\rm inter}\sigma^{yy}_{\rm inter}}$ now depends on $\Delta$ but does not diverge at the gap edge:
\begin{eqnarray}
\sigma^\Delta_{\rm inter}=&\displaystyle {e^2\over h}\sqrt{\pi\over 30}\sqrt{1+14\biggl({\Delta\over\Omega}\biggr)^2+48\biggl({\Delta\over\Omega}\biggr)^4}
\nonumber\\
&\times\theta(\Omega-2\Delta)\theta(\Omega-2\mu).
\label{eq:sigmadelta}
\end{eqnarray}
 There is also variation with frequency $\Omega$ which drops out for $\Omega>>\Delta$ and Eq.~(\ref{eq:sigmadelta}) reduces to Eq.~(\ref{eq:sigmainter}). Individually $\sigma^{xx}$ and $\sigma^{yy}$ have a dependence on $\Delta$ which goes away as $\Omega>>\Delta$. The known result for gapped graphene\cite{Gusynin:2009} with $N_f=1$ is
\begin{equation}
\displaystyle \sigma^\Delta_{\rm gr}={e^2\over h}{\pi\over 8}\biggl[1+\biggl({2\Delta\over\Omega}\biggr)^2\biggr]
\theta(\Omega-2\Delta)\theta(\Omega-2\mu).
\end{equation}
This is shown as the solid black curve in Fig.~\ref{fig6}.

For the Drude intraband contribution including an energy gap, we find
\begin{eqnarray}
\sigma_{\rm intra}^{xx}(\Omega) &= \displaystyle{e^2\over h}{12\over 5G}
{(\mu^2-\Delta^2)^{5/4}\over\mu\sqrt{ 2mv^2}} 
\delta(\Omega),\\
\sigma_{\rm intra}^{yy}(\Omega) &= \displaystyle{e^2\over h}{2\pi G\over 3}
{(\mu^2-\Delta^2)^{3/4}\over\mu} \sqrt{2mv^2}
\delta(\Omega),
\label{eq:intrabandgap}
\end{eqnarray}
which reduce to Eqs.~(\ref{eq:intrabandcleana}) and (\ref{eq:intrabandcleanb}) for $\Delta=0$.
The square root of the product of these two quantities gives
\begin{eqnarray}
\sigma^\Delta_{\rm intra}(\Omega)&=&\sqrt{\sigma^{xx}_{\rm intra}\sigma^{yy}_{\rm intra}}\\
                        &=&\displaystyle {e^2\over h}  \sqrt{8\pi\over 5}{{\mu^2-\Delta^2}\over\mu}
\delta(\Omega),
\label{eq:sqrtintragap}
\end{eqnarray}
with $\mu>\Delta$, and zero otherwise.
This properly reduces to our previous result [Eq.~(\ref{eq:sqrtintra})] when we set the gap to zero. The known result for gapped graphene\cite{Gusynin:2009} is recovered if $\sqrt{8\pi/5}$ is replaced by $\pi/2$. 

For the gapped semi-Dirac model Hamiltonian given in Eq.~(\ref{eq:Hmatrixgap}), K. Huang {\it et al.}\cite{Huang:2015} showed that because of the quadratic dependence on $k_x$ (rather than linear in pure Dirac), the Berry curvature becomes an odd function of $k_x$ (in contrast to pure Dirac  where it is even) and consequently leads to zero chern number rather than the $\pm 1$ of pure Dirac. As a result, for semi-Dirac, there is no finite Hall conductivity.

\section{Impurity scattering}

We now return to the case without an energy gap in order to discuss the effects of elastic residual impurity scattering, which particularly affects the low frequency conductivity, both widening out the intraband Drude conductivity and providing a contribution to the dc conductivity from both the intraband and interband pieces. 

So far we have evaluated the conductivity of Eq.~(\ref{eq:condformula}) only in the clean limit for which the matrix spectral density $\hat A({\bm k},\omega)$ involves Dirac $\delta$-functions and this makes it easy to evaluate Eq.~(\ref{eq:condformula}) analytically. Now we wish to consider the case when 
$\hat A({\bm k},\omega)$ 
is broadened and effects due to a self-energy $\Sigma({\bm k},\omega)$ function are considered. The components of the $2\times 2$ matrix 
$\hat A({\bm k},\omega)$ 
can be written in terms of combinations of scalar spectral densities $A_\pm({\bm k},\omega)$ given by
\begin{equation}
A_\pm({\bm k},\omega)={1\over \pi}{|{\rm Im} \Sigma({\bm k},\omega)|\over
[\omega-{\rm Re}\Sigma({\bm k},\omega)\mp \epsilon_{\bm k}]^2+[{\rm Im}\Sigma({\bm k},\omega)]^2}.
\end{equation}
The self-energy is a complex function with real and imaginary parts.

To get a first understanding of the changes that broadening can bring to the clean limit conductivity, we can replace the delta functions in the spectral density by a Lorentzian with a small constant width of $|{\rm Im}\Sigma({\bm k},\omega)|\equiv\Gamma$. In this case, the ${\rm Re}\Sigma({\bm k},\omega)$ is zero and consequently,
\begin{equation}
A_\pm({\bm k},\omega)={1\over \pi}{\Gamma\over
(\omega\mp \epsilon_{\bm k})^2+\Gamma^2}.
\label{eq:spectral}
\end{equation}
The conductivity of Eq.~(\ref{eq:condformula}) can be evaluated to read
\begin{eqnarray}
\sigma^{xx}(T,\Omega)=& \displaystyle \frac{ e^2}{h}\frac{2}{\Omega\sqrt{2mv^2}}\int^{+\infty}_{-\infty}d\omega [f(\omega) - f(\omega+\Omega)]\nonumber\\
&\times\displaystyle\int_0^\infty d\epsilon\,  {\epsilon^{3/2}}\int_0^{\pi/2}d\phi\sqrt{\cos\phi}[I_1+\cos(2\phi)I_2]\nonumber\\
& \label{eq:sigmaimpxx}\\ 
\sigma^{yy}(T,\Omega)=& \displaystyle \frac{ e^2}{h}\frac{\sqrt{2mv^2}}{2\Omega}\int^{+\infty}_{-\infty}d\omega [f(\omega) - f(\omega+\Omega)]\nonumber\\
&\times\displaystyle\int_0^\infty d\epsilon  {\sqrt{\epsilon}}\int_0^{\pi/2}d\phi{1\over\sqrt{\cos\phi}}[I_1-\cos(2\phi)I_2],\nonumber\\
& \label{eq:sigmaimpyy}
\end{eqnarray}
where 
\begin{equation}
I_1= [A_+(\omega)+A_-(\omega)][A_+(\omega+\Omega)+A_-(\omega+\Omega)]
\end{equation}
and
\begin{equation}
I_2= [A_+(\omega)-A_-(\omega)][A_+(\omega+\Omega)-A_-(\omega+\Omega)],
\end{equation}
and for convenience we have suppressed the ${\bm k}$ labels on the spectral density, or equivalently, as used here, the phase $\phi$ and the energy $\epsilon$ (see the Appendix). Note that Eqs.~(\ref{eq:sigmaimpxx}) and (\ref{eq:sigmaimpyy}) contain both interband [$A_{\pm}(\omega)A_{\mp}(\omega+\Omega)$] 
and intraband [$A_{\pm}(\omega)A_{\pm}(\omega+\Omega)$] optical transitions. 
In our model of constant $\Gamma$, the conductivity obeys a scaling behavior:
\begin{eqnarray}
\sigma^{xx}(T,\mu,\Omega)&\equiv \displaystyle{e^2\over h}\sqrt{\Gamma\over mv^2}
\bar\sigma^{xx}(\bar T,\bar \mu, \bar\Omega)\label{eq:xxsigimp}\\
\sigma^{yy}(T,\mu,\Omega)&\equiv \displaystyle{e^2\over h}\sqrt{mv^2\over\Gamma}
\bar\sigma^{yy}(\bar T,\bar \mu, \bar\Omega),
\label{eq:yysigimp}
\end{eqnarray}
where any variable $\bar x\equiv x/\Gamma$.

\begin{figure}
\includegraphics[width=0.9\linewidth]{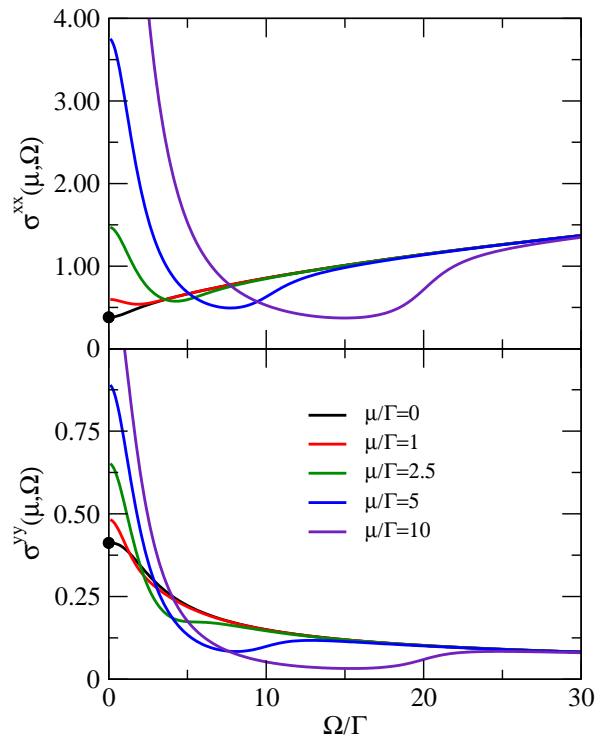}
\caption{The real part of the conductivity 
as a function of the normalized photon energy $\Omega/\Gamma$. Here, impurity scattering is included through the parameter $\Gamma$ and a set of curves is plotted for
varying $\mu/\Gamma$. The upper frame is for the case of $\sigma^{xx}(\Omega)$ normalized to $(e^2/h)\sqrt{mv^2/\Gamma}$ and the lower frame is for  $\sigma^{yy}(\Omega)$ normalized to $(e^2/h)\sqrt{\Gamma/(mv^2)}$.
}\label{fig7}
\end{figure}

For $T=0$, the two functions $\bar\sigma^{xx}$ and $\bar\sigma^{yy}$ are universal functions of $\bar\mu$ and $\bar\Omega$ which can be plotted as a family of curves with label $\bar\mu$ as a function of $\bar\Omega$. In Fig.~\ref{fig7}, we show our numerical results for $\bar\mu=0$, 1, 2.5, 5 and 10, evaluated from
Eqs.~(\ref{eq:xxsigimp}) and (\ref{eq:yysigimp}). The top frame is for $\bar\sigma^{xx}(\mu,\Omega)$ or
$\sigma^{xx}(\mu,\Omega)$ normalized by ${e^2\over h}\sqrt{\Gamma\over mv^2}$
and the bottom frame is for $\bar\sigma^{yy}(\mu,\Omega)$ or
$\sigma^{yy}(\mu,\Omega)$ in units of 
${e^2\over h}\sqrt{mv^2\over\Gamma}$. The solid black curve in the top frame is for zero doping (charge neutrality) and shows features of the $\sqrt{\Omega}$ dependence of the clean limit (solid blue curve of Fig.~\ref{fig2}). The main effect of including a finite residual scattering rate is to make $\sigma^{xx}(\mu,\Omega)$ finite at $\Omega=0$ rather than going to zero as in the clean limit case. Later we will provide an analytic formula for the value of the dc conductivity at charge neutrality which agrees perfectly with the numerical results presented here (heavy solid black point). We note that because $\mu=0$, there is no Drude contribution. By contrast, all other curves show a broadened Drude peak about $\Omega=0$, which increases in magnitude as $\mu$ is increased, and more optical spectral weight resides in the intraband transition which increases as $\mu^{3/2}$ in the clean limit [see Eq.~(\ref{eq:intrabandcleana})] We also see a depression of the interband background following the Drude peak before rising again around $\Omega=2\mu$ to return to the clean limit interband background value. In the lower frame of Fig.~\ref{fig7}, we observe very much the same trends as described for  $\sigma^{xx}$. In the solid black curve which applies at charge neutrality, we see the $1/\sqrt{\Omega}$ behavior of the clean limit (solid red curve of Fig.~\ref{fig2}) except that the singularity at $\Omega\to 0$ is removed when $\Gamma\ne 0$ and the dc conductivity becomes finite (heavy solid black dot, for emphasis).

\begin{figure}
\includegraphics[width=0.9\linewidth]{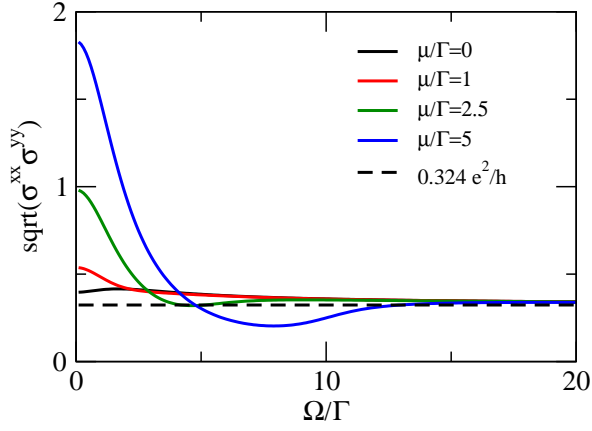}
\caption{The real part of the defined interband conductivity $\sigma=\sqrt{\sigma^{xx}\sigma^{yy}}$ in units of $e^2/h$
as a function of the photon energy $\Omega$ normalized to the impurity parameter $\Gamma$. Curves are plotted for various $\mu/\Gamma$. The black dashed curve is the case for  the universal background of the semi-Dirac case with $\mu=\Gamma=0$.
}\label{fig8}
\end{figure}

Fig.~\ref{fig8} presents results for the square root of the product $\sigma^{xx}(\mu,\Omega)\sigma^{yy}(\mu,\Omega)$. In this case, the material parameters $m$ and $v$ have dropped out as has the scattering rate $\sqrt{\Gamma}$ which appears in Eqs.~(\ref{eq:xxsigimp}) and (\ref{eq:yysigimp}) so that the remaining prefactor unit is simply $e^2/h$. There are only small differences between these curves and the corresponding curves for graphene (not shown). One difference seen in the case of $\mu=0$ (black curve for charge neutrality) is a slight increase above the universal semi-Dirac background (black dashed curve) in the region shown which is not seen in graphene. This effect is perhaps not unexpected  since in semi-Dirac $\sigma^{yy}$ diverges as $1/\sqrt{\Omega}$ (red curve of Fig.~\ref{fig2}) in the clean limit. A further difference is that the value of the saturated background at large photon energy is slightly different in semi-Dirac compared with graphene.

Next we consider the dc limit of the conductivity at charge neutrality ($\mu\to 0$) and zero temperature ($T\to 0$). Eqs.~(\ref{eq:sigmaimpxx}) and
(\ref{eq:sigmaimpyy}) reduce to
\begin{eqnarray}
\sigma^{xx}_{\rm dc}=& \displaystyle \frac{ e^2}{h}\frac{2}{\sqrt{2mv^2}}{1\over G}\int^{+\infty}_{0}d\epsilon  \epsilon^{3/2}[A_+(0)+A_-(0)]^2\\
\sigma^{yy}_{\rm dc}=& \displaystyle \frac{ e^2}{h}\frac{\sqrt{2mv^2}}{2}
\pi G\int_0^\infty d\epsilon  \sqrt{\epsilon}[A_+(0)+A_-(0)]^2,
\end{eqnarray}
with the reminder that  the $A$'s depend on $\epsilon$.
The integration over $\epsilon$ can be done noting that:
\begin{equation}
\int^\infty_0\displaystyle{\epsilon^{1/2}\over (1+\epsilon^2)^2}d\epsilon=\int^\infty_0\displaystyle{\epsilon^{3/2}\over (1+\epsilon^2)^2}d\epsilon
={\pi\over 4\sqrt{2}}
\end{equation}
to obtain
\begin{eqnarray}
\sigma^{xx}_{\rm dc}=& \displaystyle \frac{ e^2}{h}\sqrt{{\Gamma\over mv^2}}{1.20\over\pi}\label{eq:dc1}\\
\sigma^{yy}_{\rm dc}=& \displaystyle \frac{ e^2}{h}\sqrt{{mv^2\over\Gamma}}{1.31\over\pi}.\label{eq:dc2}
\end{eqnarray}
The square root of their product 
$\sigma_{dc}\equiv\sqrt{\sigma^{xx}_{\rm dc}\sigma^{yy}_{\rm dc}}$ 
is $1.25e^2/(\pi h)$. These numbers for the dc conductivity at charge neutrality agree well with our numerical calculations based on Eqs.~(\ref{eq:sigmaimpxx})-(\ref{eq:sigmaimpyy}) which are plotted in Fig.~\ref{fig7}. The solid dots on the vertical axis are from our analytic results mentioned above. This provides a check on both our numerical and analytical work.

It is of interest to compare our results for the dc limit at charge neutrality with our previous results for the value of the interband background associated with the quantity $\sigma=\sqrt{\sigma^{xx}\sigma^{yy}}$. For graphene (with $N_f=1$), $\sigma_{\rm inter}=\pi e^2/(8h)$ and $\sigma_{\rm dc}=e^2/(\pi h)$, which gives $\sigma_{\rm dc}/\sigma_{\rm inter}=8/\pi^2=0.811$. For semi-Dirac,  $\sigma_{\rm inter}=0.324 e^2/h$ and $\sigma_{\rm dc}=0.397 e^2/h$, with $\sigma_{\rm dc}/\sigma_{\rm inter}=1.22$ which  is $\approx 3/2$ times the graphene value. Universal limits for transport (independent of impurity scattering) have been discussed previously in other contexts, for example, in d-wave superconductors\cite{Lee:1993}. 

\begin{figure}
\includegraphics[width=0.9\linewidth]{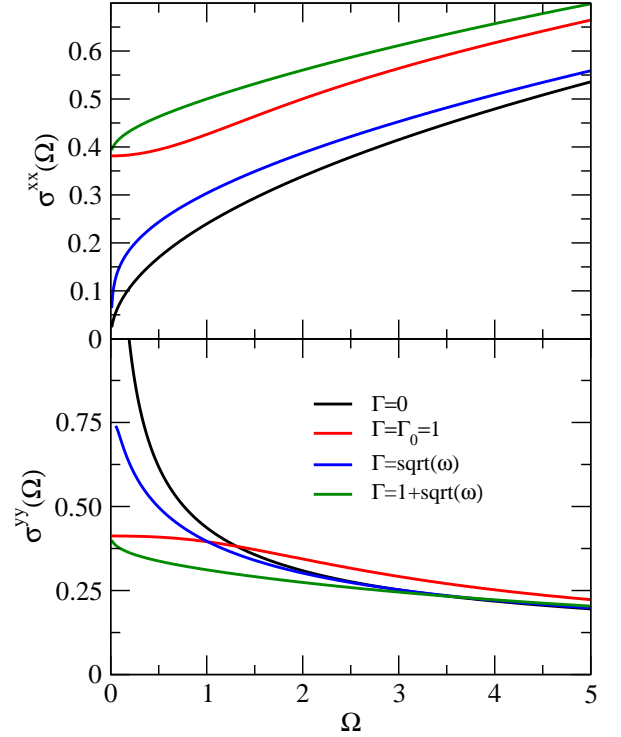}
\caption{The interband conductivity $\sigma^{xx}_{\rm inter}(\Omega)$ (top frame)
and $\sigma^{yy}_{\rm inter}(\Omega)$ (bottom frame) in units of $e^2/h$
as a function of photon energy in units of $mv^2$. The black curves are in the clean limit ($\Gamma = 0$) while the red curves include a constant scattering rate of  $\Gamma=\Gamma_0=1$ in units of $mv^2$. The blue curves include an energy dependence in the scattering rate of the form $\Gamma=\sqrt{\omega}$ and the green curves have $\Gamma=1+\sqrt{\omega}$.
}\label{fig9}
\end{figure}

One could go beyond a constant $\Gamma$ approximation and treat residual scattering in a self-consistent Born approximation, for instance, based on a specific model for the impurity scattering. Or more simply, as an illustrative model, $\Gamma$ could be taken to be proportional to the density of states $N(\epsilon)\sim\sqrt{\epsilon}$ for semi-Dirac. Any model scattering with $|{\rm Im}\Sigma(\omega)|\sim \omega^l$ can be treated as an approximate phenomenological model\cite{Lundgren:2014,Tabert:2016b} ignoring the corresponding real part of the self-energy which leads to a shift in quasiparticle energy. This is easily incorporated in our numerical calculations by including a frequency dependence in the  $\Gamma$ in the $A$'s with ${\rm Re}\Sigma(\omega)$ set equal to zero for simplicity. Results for the conductivity are given in Fig.~\ref{fig9}. The black curves in Fig.~\ref{fig9} are for comparison and are in the clean limit based on Eqs.(\ref{eq:intercleanxx}) and (\ref{eq:intercleanyy}), respectively, for $\sigma^{xx}_{\rm inter}(\Omega)$ (top)
and $\sigma^{yy}_{\rm inter}(\Omega)$ (bottom) in units of $e^2/h$. The photon energy $\Omega$ is in units of $mv^2$. 
For $\Omega\to 0$, the top curve goes to zero as $\sqrt{\Omega}$ while the bottom curve diverges as $1/\sqrt{\Omega}$. The red curves include a constant residual scattering $\Gamma=1$ in units of $mv^2$. We note that the impurity scattering pushes the curve up as compared with the clean limit for $\sigma^{xx}_{\rm inter}(\Omega)$ while it pushes it down for
$\sigma^{yy}_{\rm inter}(\Omega)$ in the small photon energy region. The blue curves are for an energy dependent scattering rate of the form $\Gamma=\sqrt{\omega}$. This behavior results if it is assumed to be directly dependent on the electronic density of states which in the semi-Dirac case has a square root dependence. As expected, in the top frame the blue curve is above the black but merges with it at $\Omega=0$ while in the bottom frame it is below the clean limit but diverges as $\Omega\to 0$. The green curves show the result when the model scattering rate is taken to be of the form $\Gamma=1+\sqrt{\omega}$. In this case, the green curve is above the red but merges with it at $\Omega=0$ in the top frame while in the bottom frame, it is below. It is clear that different models for the energy dependence of the residual scattering rate can modify the energy dependence of the conductivity but there are no significant qualitative changes.

 Another issue is possible anisotropy in the residual scattering. Adroguer {\it et. al.}\cite{Adroguer:2016} have found within the diffusive regime, in semi-Dirac, that the Fermi surface anisotropy and the nature of the eigenstates can lead to significant anisotropy in the resulting residual scattering rate. In our previous analysis, we have assumed that $\Gamma$ was constant, but the transformation from $\epsilon$ to $\bar \epsilon=\epsilon/\Gamma$ which we have used can still be done even if $\Gamma$ is dependent on angle $\phi$. For the case studied here, Adrouger {\it et al.}\cite{Adroguer:2016} find that $\Gamma(\phi)$ takes the form $\Gamma_0[1+n_0\cos\phi]$ with $n_0\simeq 0.457$. Including this in our analysis, we arrive at the generalizations 
\begin{eqnarray}
\sigma^{xx}_{\rm dc}=& \displaystyle \frac{ e^2}{h}\sqrt{{\Gamma_0\over mv^2}}{1\over\pi}\int_0^{\pi/2}\sqrt{\cos\phi}[1+0.457\cos\phi]^{1/2}d\phi\nonumber\\ &\\
\sigma^{yy}_{\rm dc}=& \displaystyle \frac{ e^2}{h}\sqrt{{mv^2\over\Gamma_0}}{1\over\pi}\int_0^{\pi/2}{1\over 2\sqrt{\cos\phi}}[1+0.457\cos\phi]^{-1/2}d\phi,\nonumber\\
& 
\end{eqnarray}
which correspond to a change of constants 1.20 and 1.31 in Eqs.~({\ref{eq:dc1}) and ({\ref{eq:dc2}) to new constants equal to 1.38 and 1.20, respectively. The anisotropy in the scattering rate has increased the value of $\sigma^{xx}_{\rm dc}$ and decreased $\sigma^{yy}_{\rm dc}$. These changes are of order of 10\%.

\section{Generalization to other dispersion curves beyond semi-Dirac}

We now consider an electronic dispersion curve arising from a Hamiltonian of the general form
\begin{equation}
\hat H=\left(\begin{array}{cc}
0 & c_xk_x^n-ic_yk_y^s\\
\displaystyle c_xk_x^n+ic_yk_y^s & 0
\end{array}\right) ,
\label{eq:Hmatrixgeneral}
\end{equation}
with $n$ and $s$ integers and
$c_x$ and $c_y$ are two material-dependent coefficients. We can derive general results for the conductivity in terms of these parameters. The resulting dispersion curve will be $\epsilon_{\bm k}=\pm\sqrt{c_x^2k_x^{2n}+c_y^2k_y^{2s}}$, with $c_x=c_y=\hbar v$ and $n=s=1$ yielding the graphene Dirac case, and $(n,s)=(2,1)$ and $c_x\neq c_y$ being semi-Dirac in character.
The interband conductivities become
\begin{eqnarray}
\sigma^{xx}_{\rm inter}(\Omega)=& \displaystyle {e^2\over 4h}{n\over s}\biggl[{\Omega\over 2c_x}\biggr]^{-{1\over n}}
\biggl[{\Omega\over 2c_y}\biggr]^{{1\over s}} C_{s,n}\theta(\Omega-2\mu),
\label{eq:genxx}\\
\sigma^{yy}_{\rm inter}(\Omega)=& \displaystyle {e^2\over 4h}{s\over n}\biggl[{\Omega\over 2c_x}\biggr]^{{1\over n}}\biggl[{\Omega\over 2c_y}\biggr]^{-{1\over s}} C_{n,s}\theta(\Omega-2\mu),\label{eq:genyy}
\end{eqnarray}
with
\begin{equation}
C_{n,s} = 2\int_0^{\pi/2} [\cos\theta]^{({1\over n}+1)}[\sin\theta]^{(1-{1\over s})}d\theta.
\end{equation}
For the semi-Dirac case primarily featured in this paper, these equations properly reduce to Eqs.~(\ref{eq:intercleanxx}) and (\ref{eq:intercleanyy}) when $(n,s)=(2,1)$ and $c_x=\hbar^2/2m$ and $c_y=\hbar v$.

For the square root of the product, $\sigma_{\rm inter}=\sqrt{\sigma^{xx}_{\rm inter}\sigma^{yy}_{\rm inter}}$, we obtain
\begin{equation}
\sigma_{\rm inter}=\displaystyle {e^2\over 4h}\sqrt{C_{n,s}C_{s,n}}\,\theta(\Omega-2\mu),
\end{equation}
which is constant independent of photon energy.

If we take $n=s$ and $c_x=c_y$ which neglects anisotropy in the electronic dispersion curve, we recover the result of B\'acsi and Virosztex\cite{Bacsi:2013} with $n$ not necessarily integral. If, however, anisotropy in the $(c_x,c_y)$ is accounted for,  $\sigma^{xx}$ and $\sigma^{yy}$ differ by a factor which is $(1/c_x)^{-1/n}(1/c_y)^{1/s}$ for the $xx$ direction and $(1/c_x)^{1/n}(1/c_y)^{-1/s}$ for the $yy$ direction. Again, there is no $\Omega$ dependence. The optical response in fact is independent of photon energy $\Omega$ in all cases provided only that $n=s$. It will have dependence on photon energy in all other cases for which $xx$ goes like $\Omega^{{1\over s}-{1\over n}}$ and $yy$ goes like the inverse dependence $\Omega^{{1\over n}-{1\over s}}$. 
This implies the $\Omega$ will drop out of the product of $\sigma^{xx}$ and $\sigma^{yy}$ so that
$\sqrt{\sigma^{xx}\sigma^{yy}}$ is still constant.

As $n$ is increased, the dispersion curves flatten out at low energy and this leads to a change in the $C_{n,s}$ coefficients even for $s=n$. 
The $C_{1,1}$ corresponds to graphene and $C_{1,2}=C_{2,1}$ to semi-Dirac. In these limits, we recover the result that the constant interband background for graphene is 0.393 and for semi-Dirac 0.324 in units of $e^2/h$. In general, this background will vary with value of $n$ and $s$ and can be greater as well as smaller than for graphene. 

Next we consider the density of states in our anisotropic model.
It is given by
\begin{eqnarray}
N(\omega)&=&\displaystyle\int {d^2k\over (2\pi)^2}\delta(\omega-\epsilon_{\bm k})\nonumber\\
&=&\displaystyle
\bigg({1\over c_x}\biggr)^{1\over n}\bigg({1\over c_y}\biggr)^{1\over s}\omega^{({1\over n}+{1\over s}-1)}{D_{n,s}\over ns\pi^2}
\end{eqnarray}
with
\begin{equation}
D_{n,s}=\int_0^{\pi/2}d\theta[\cos\theta]^{({1\over n}-1)}[\sin\theta]^{({1\over s}-1)}.
\end{equation}
In the case $n=s$, which according to  Eqs.(\ref{eq:genxx}) and (\ref{eq:genyy}) corresponds to the $\Omega$-independent constant
$\sigma^{xx}_{\rm inter}$ and $\sigma^{yy}_{\rm inter}$, the density of states takes the form
\begin{equation}
N(\omega)=\displaystyle
\bigg({1\over c_xc_y}\biggr)^{1\over n}\omega^{({2\over n}-1)}{D_{n,n}\over n^2\pi^2},
\end{equation}
which is a constant independent of energy $\omega$ only for  the case $n=2$. For $n=1$, we recover the well-known result for graphene, namely, $N(\omega)=\omega/(2\pi^2v^2)$, with $c_x=c_y=v$ ($\hbar=1$). For $n=2$, we get $N(\omega)$ constant and equal to $1.85m/\pi^2$, which is different in magnitude from the density of states of a two-dimensional electron gas with quadratic dispersion which is $m/2\pi$. For the case of semi-Dirac with $c_x=1/2m$ and $c_y=v$, we recover the known result\cite{Banerjee:2012} $N(\omega)=1.31\sqrt{2m\omega}/(v\pi^2)$, which varies like the square root of energy. For a general $(n,s)$, it is interesting to compare the energy dependence of the conductivity background with that of the density of states. We find $N(\Omega)/\sigma^{xx}_{\rm inter}(\Omega)\sim \Omega^{[(2/n)-1]}$ 
and $N(\Omega)/\sigma^{yy}_{\rm inter}(\Omega)\sim \Omega^{[(2/s)-1]}$ which always depends on energy $\Omega$ provided $s$ and $n$ are different from $2$. 

It is also of interest to find the dc conductivity at charge neutrality when the  Hamiltonian of Eq.~(\ref{eq:Hmatrixgeneral}) is employed. It is given by
\begin{eqnarray}
\sigma^{xx}_{\rm dc}=& \displaystyle \frac{ e^2}{4h}\int d^2k v_x^2
[A_+(0)+A_-(0)]^2,\label{eq:gendcxx}\\
\sigma^{yy}_{\rm dc}=& \displaystyle \frac{ e^2}{4h}\int d^2k v_y^2
[A_+(0)+A_-(0)]^2.\label{eq:gendcyy}
\end{eqnarray}
Here, $v_x= c_xnk_x^{n-1}$ and $v_y=c_ysk_y^{s-1}$ are the velocities. After considerable but straightforward algebra Eqs.~(\ref{eq:gendcxx}) and (\ref{eq:gendcyy})
can be reduced to 
\begin{eqnarray}
\sigma^{xx}_{\rm dc}=& \displaystyle \frac{ 4e^2n}{\pi^2hs}\biggl({1\over c_x}\biggr)^{-{1\over n}}\biggl({1\over c_y}\biggr)^{1\over s}\Gamma^{({1\over s}-{1\over n})}H_{s,n}J_{s,n},\\
\sigma^{yy}_{\rm dc}
=& \displaystyle \frac{ 4e^2s}{\pi^2hn}\biggl({1\over c_x}\biggr)^{1\over n}\biggl({1\over c_y}\biggr)^{-{1\over s}}\Gamma^{({1\over n}-{1\over s})}H_{n,s}J_{n,s},
\end{eqnarray}
where 
\begin{equation}
H_{n,s}=\int_0^{\pi/2}d\theta[\cos\theta]^{({1\over n}-1)}[\sin\theta]^{(1-{1\over s})}.
\end{equation}
and
\begin{equation}
J_{n,s}=\displaystyle\int_0^{\infty}dx{x^{({1\over n}-{1\over s}+1)}\over (1+x^2)^2}.
\end{equation}
Also,
\begin{equation}
\sqrt{\sigma^{xx}_{\rm dc}\sigma^{yy}_{\rm dc}}={4e^2\over \pi^2 h}\sqrt{H_{n,s}H_{s,n}J_{n,s}J_{s,n}}.
\end{equation}
Importantly, we note that the residual quasiparticle scattering rate $\Gamma$ has dropped out of this quantity which is, in that sense, universal but still dependent on $n$ and $s$.
For semi-Dirac, $c_x=1/(2m)$ and $c_y=v$, $n=2$ and $s=1$, and we recover the result discussed in a previous section and for $n=s=1$, $c_x=c_y=v$, we find the known result for graphene, {\it i.e.} $e^2/(\pi h)$, when the degeneracy factor $N_f=4$ used in the graphene literature is left out. In general, for different choices  of $n$ and $s$ there is some variation of the magnitude of the ``universal'' dc limit of the conductivity at charge neutrality.

\begin{figure}
\includegraphics[width=0.9\linewidth]{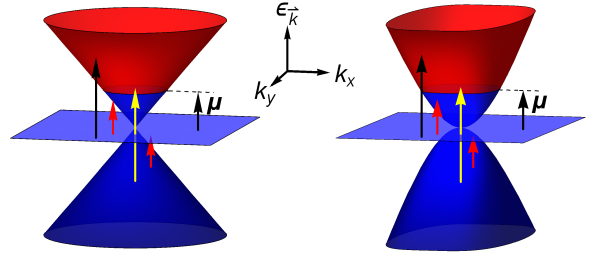}
\caption{Left: the energy dispersion in the fully Dirac $\alpha$-${\cal T}_3$ model exhibiting both Dirac cones and a flat band at zero energy. Right: the semi-Dirac version of the $\alpha$-${\cal T}_3$ model. 
Red and blue represent unoccupied and occupied states, respectively.  The arrows indicate typical transitions: black for absorptive, red for Pauli-blocked, and yellow for blocked by selection rules if $\alpha=1$ but otherwise absorptive for $\alpha<1$.
}\label{fig10}
\end{figure}

\section{The semi-Dirac $\alpha$-${\cal T}_3$ model}

Graphene is often referred to as a pseudospin-1/2 system in reference to the spin-1/2 type Pauli matrices which are used to define a low energy Hamiltonian in relation to the two triangular sublattices, {\it i.e.} $\hat H=\hbar v {\bm k}\cdot {\bm \sigma}$. Various other models related to general pseudospin $S$ have also been examined with an emphasis on the optical conductivity.\cite{Dora:2011} The pseudospin 1 version is also known as the dice lattice or the ${\cal T}_3$ lattice and uses the $S=1$ matrices in $\hat H=\hbar v {\bm k}\cdot {\bm S}$:
\begin{equation}
\hat{H} = \left( \begin{matrix}
0 & f_{\bm k} & 0 \\
f^{*}_{\bm k} & 0 & f_{\bm k} \\
0 & f^{*}_{\bm k} & 0 \\
\end{matrix} \right),
\end{equation}
where  $f_{\bm k}=\hbar v k_x-i\hbar vk_y$. This Hamiltonian gives rise to two linear energy dispersions or cones, as in graphene, plus a flat band at charge neutrality: $\epsilon_{\bm k}=0, \pm\hbar v|{\bm k}|$ (see Fig.~\ref{fig10}, left side). The origin of this model is in the low energy limit of a tight-binding model which has a hopping from A and B sublattices forming a honeycomb lattice as in graphene, but now including a central site in the middle of the hexagon (a new sublattice labelled C) which is only coupled to the A sublattice, for instance, but with same strength of hopping as the A to B. The optical conductivity for this model and the extension to general $S$ has been given in Ref.~{\onlinecite{Dora:2011}}. An interesting variation on this model has been the $\alpha$-${\cal T}_3$ model which introduces a variable parameter $\alpha$ that multiplies the hopping parameter from A to C resulting in a Hamiltonian of the form:
\begin{align}
\hat{H} &= \left( \begin{matrix}
0 & c_\alpha f_{\bm k} & 0 \\
c_\alpha f^{*}_{\bm k} & 0 & s_\alpha f_{\bm k} \\
0 & s_\alpha f^{*}_{\bm k} & 0 \\
\end{matrix} \right) \, ,
\label{eq:s1}
\end{align}
where
\begin{align}
c_{\alpha} &= \frac{1}{\sqrt{1+\alpha^{2}}} \, , \\
s_{\alpha} &= \frac{\alpha}{\sqrt{1+\alpha^{2}}} \, .
\end{align}
The Hamiltonian represents a superposition of the $S=1/2$ and $S=1$, where $\alpha=0$ reduces to the graphene case and $\alpha=1$ gives the dice or ${\cal T}_3$ case. Intermediate to these two cases, there is a variable Berry phase. Evidence for the intermediate type behaviour has been presented in the work of Malcolm and Nicol\cite{Malcolm:2015} upon examination of the results of magneto-optics measurements on a Hg-Cd-Te type of material. The conductivity in zero magnetic field of the $\alpha$-${\cal T}_3$ model has been calculated by Illes {\it et al.}\cite{Illes:2015}.

Here, we consider a semi-Dirac form of this Hamiltonian, which has been proposed by Piechon \textit{et al.}\cite{Piechon:2015}. The Hamiltonian is of the same form as Eq.~(\ref{eq:s1}), but the semi-Dirac form is used for $f_{\bm k}$:
\begin{widetext} 
\begin{align}
\hat{H} &= \displaystyle
\left( \begin{matrix}
\displaystyle 0 & \displaystyle c_{\alpha} ({\hbar^2 k_x^2\over 2m}-i\hbar v k_y) & 0 \\
\displaystyle c_{\alpha} ({\hbar^2k_x^2\over 2m}+i\hbar v k_y) & 0 & \displaystyle
s_{\alpha} ({\hbar^2k_x^2\over 2m}-i\hbar v k_y) \\
\displaystyle 0 & \displaystyle s_{\alpha} ({\hbar^2k_x^2\over 2m}+i\hbar v k_y) & 0\\
\end{matrix} \right) .
\end{align}
\end{widetext}
The energy dispersion is given as three bands: $\epsilon_{\bm k}=0$, $\pm\sqrt{[\hbar^2k_x^2/(2m)]^2+(\hbar v k_y)^2}$ and is shown on the right side of Fig.~\ref{fig10}.
We have used this Hamiltonian to evaluate the conductivity from the Kubo formula previously discussed. For use in numerical work, that could allow for the effects of impurity scattering or other self-energy effects, the conductivity can be evaluated for $T=0$ from
\begin{widetext}
\begin{align}
\sigma^{x x} (\Omega) &= \displaystyle\frac{N_f e^2}{h}\frac{8}{5G\sqrt{2mv^2}}\frac{1}{\Omega} \int^{\mu}_{\mu - \Omega} \mathop{d \omega} \int_{0}^{\infty} \epsilon^{3/2} \mathop{d \epsilon} \, \Bigg[ \frac{3}{2}I_{\rm intra}+ {(\alpha^2-1)^2\over(\alpha^2+1)^2}I_{\rm inter,cones}+{2\alpha^2\over (\alpha^2+1)^2}I_{\rm inter,flat}
\Bigg],\label{eq:kylexx} \\
\sigma^{yy} (\Omega) &=\displaystyle \frac{N_f e^2}{h}{\pi G\sqrt{2mv^2}\over 3}\frac{1}{\Omega} \int^{\mu}_{\mu - \Omega} \mathop{d \omega} \int_{0}^{\infty} \epsilon^{1/2} \mathop{d \epsilon} \, \Bigg[ 2I_{\rm intra}+ {(\alpha^2-1)^2\over(\alpha^2+1)^2}I_{\rm inter,cones}+{2\alpha^2\over (\alpha^2+1)^2}I_{\rm inter,flat}\Bigg], \label{eq:kyleyy}
\end{align} 
\end{widetext}
where
\begin{eqnarray}
I_{\rm intra}=& A_+(\omega)A_+(\omega+\Omega)+A_-(\omega)A_-(\omega+\Omega),\nonumber\\
&\label{eq:Iintra}\\
I_{\rm inter,cones}=& A_+(\omega)A_-(\omega+\Omega)+A_-(\omega)A_+(\omega+\Omega),\nonumber\\
&\\
I_{\rm inter,flat}=& A_0(\omega)A_-(\omega+\Omega)+A_0(\omega)A_+(\omega+\Omega)\nonumber
\\
&+A_0(\omega+\Omega)A_-(\omega)+A_0(\omega+\Omega)A_+(\omega).\nonumber\\
&
\end{eqnarray}
The $A$'s above have the same form as in Eq.~(\ref{eq:spectral}) with the subscript $0$ referring to $\epsilon_{\bm k}=0$ (the flat band) and the $\pm$ referring to the upper and lower dispersions, respectively.  The subscript label ``inter,cones'' refers to interband transitions between the upper and lower bands that, while now modified in semi-Dirac, would be cones in the Dirac version of $\alpha$-${\cal T}_3$. The label ``inter,flat'' refers to interband transitions involving the flat band, between it and either of the upper or lower band.

These equations can be reduced further to simple analytical results in the clean limit ($\Gamma=0$). The intraband terms are found to be (taking $N_f=1$):
\begin{eqnarray}
\sigma_{\rm intra}^{xx}(\Omega) &=& \displaystyle{e^2\over h}{12\over 5G}{\mu^{3/2}\over\sqrt{2mv^2}} \delta(\Omega),\label{eq:at3intraxx}
\\
\sigma_{\rm intra}^{yy}(\Omega) &=& \displaystyle{e^2\over h}{2\pi G\over 3}\sqrt{2mv^2\mu}\delta(\Omega).
\label{eq:at3intrayy}
\end{eqnarray}
We note that they do not depend on the parameter $\alpha$ as was also seen in the pure Dirac version of the $\alpha$-${\cal T}_3$ model. Indeed, they are the same as Eqs.~(\ref{eq:intrabandcleana}) and (\ref{eq:intrabandcleanb}) for the semi-Dirac case, as the intraband terms are solely associated with the semi-Dirac cones, see Eq.~(\ref{eq:Iintra}).
The interband terms do depend on $\alpha$:
\begin{widetext}
\begin{eqnarray}
\sigma_{\rm inter}^{xx}(\Omega) &=& \displaystyle{e^2\over h}{16\over 5G}\sqrt{\Omega\over  2mv^2}
\biggl[{\alpha^2\over (\alpha^2+1)^2}\theta(\Omega-\mu)+{1\over 8\sqrt{2}}
{(\alpha^2-1)^2\over(\alpha^2+1)^2}
\theta(\Omega-2\mu)\biggr]
,\label{eq:at3interxx}
\\
\sigma_{\rm inter}^{yy}(\Omega) &=& \displaystyle{e^2\over h}{2\pi G\over 3}\sqrt{2 mv^2\over \Omega}
\biggl[
{\alpha^2\over (\alpha^2+1)^2}
\theta(\Omega-\mu)+{1\over 4\sqrt{2}}{(\alpha^2-1)^2\over(\alpha^2+1)^2}\theta(\Omega-2\mu)\biggr].
\label{eq:at3interyy}
\end{eqnarray}
\end{widetext}
For $\alpha=0$, we recover Eqs.~(\ref{eq:intercleanxx}) and (\ref{eq:intercleanyy}) as expected. The interband absorption is entirely from intercone transitions which start at $\Omega=2\mu$. For $\alpha=1$, we find the result for the semi-Dirac version of the $S=1$ dice of $\alpha$-${\cal T}_3$ lattice, where selection rules dictate that transitions between the semi-Dirac cones cannot occur and only transitions from the flat band to the cones can happen (see Fig.~\ref{fig10}). In this case, the interband absorption edge starts at $\Omega=\mu$. For $0<\alpha<1$, there is an admixture of the two $S=1/2$ and $S=1$ behaviours as previously discussed which can be related to a variable Berry phase.\cite{Illes:2015} 

\begin{figure}
\includegraphics[width=0.9\linewidth]{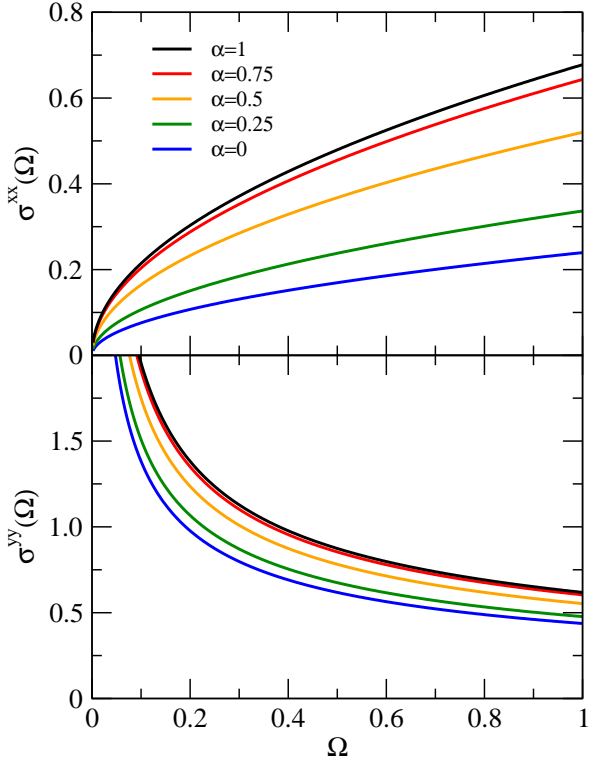}
\caption{The optical conductivity $\sigma^{xx}(\Omega)$
and $\sigma^{yy}(\Omega)$ for $\mu=0$ and for varying $\alpha$ as shown in the figure. The conductivity is in units of $e^2/h$ and $\Omega$ is normalized by $mv^2$.  
}\label{fig11}
\end{figure}

In Fig.~\ref{fig11}, we show the results for $\mu=0$ 
which illustrates the interband conductivity with no absorption edges produced by Pauli-blocking which arises with finite $\mu$. In this case, the $\sqrt{\Omega}$ and $1/\sqrt{\Omega}$ behaviour of the $\sigma^{xx}(\Omega)$
and $\sigma^{yy}(\Omega)$, respectively, are seen very clearly. We note that the overall amplitude of this behaviour is modified by the $\alpha$ parameter as shown, with a monotonic progression to greatest amplitude at $\alpha=1$. The amplitude factor is slightly different in the $xx$ versus the $yy$ directions as can be seen by examining the coefficients in front of the theta factors in 
Eqs.~(\ref{eq:at3interxx}) and  (\ref{eq:at3interyy}) which add together here for $\mu=0$ (or in general when $\Omega>2\mu$). The variation in $xx$ is
given by $[8\sqrt{2}\alpha^2+(\alpha^2-1)^2]/(\alpha^2+1)^2$
and in $yy$ by $[4\sqrt{2}\alpha^2+(\alpha^2-1)^2]/(\alpha^2+1)^2$.

\begin{figure}
\includegraphics[width=0.9\linewidth]{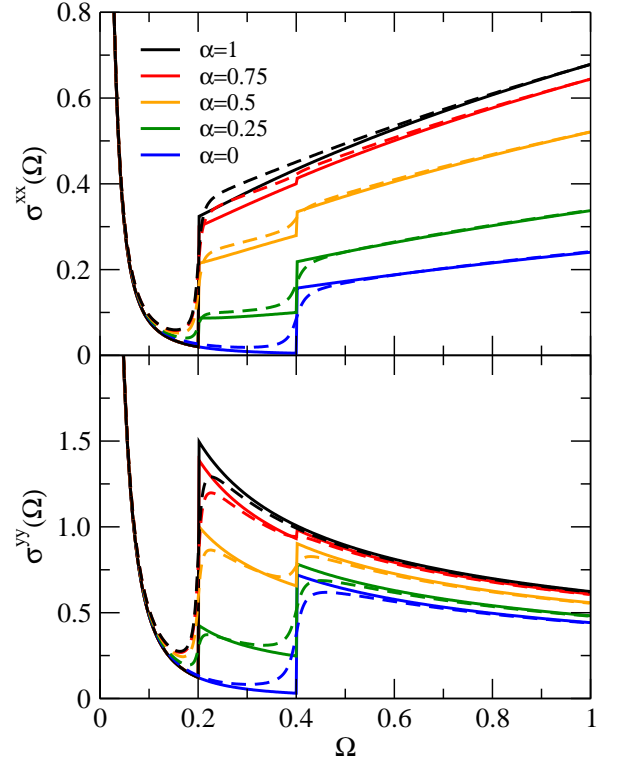}
\caption{The optical conductivity $\sigma^{xx}(\Omega)$
and $\sigma^{yy}(\Omega)$ for $\mu=0.2$ and for varying $\alpha$ as shown in the figure. The solid curves are made using the analytic formulas with a broadened Drude as described in the text and the dashed curves are numerical calculations using Eqs.~(\ref{eq:kylexx}) and (\ref{eq:kyleyy}) with $\Gamma=0.007$. 
The quantities $\Omega$ and $\Gamma$ are normalized by $mv^2$ and the conductivity is normalized $e^2/h$.
}\label{fig12}
\end{figure}

In Fig.~\ref{fig12}, we show the conductivity with finite $\mu=0.2$ where now a double step appears for $0<\mu<1$. For $\alpha=0$, the result is the usual semi-Dirac case discussed in the beginning of this paper with an absorption edge at $2\mu$ for interband processes and a low frequency peak for the Drude (intraband result). For $\alpha=1$, the conductivity also has a single absorption edge at $\mu$ for the flat band to cone transitions. Intermediate $\alpha$ corresponds to the admixture of the two limits. What is notable about these results is the signature that is shown in the interband conductivity of two steps which would be a feature of the $\alpha$-${\cal T}_3$ behaviour. At this level, it is harder to see the square root dependence in the $xx$ curves but they remain robust. The underlying Drude conductivity peak can change this behaviour somewhat as the Drude peak is finite but decays with  increasing $\Omega$. Consequently the green curve in the $xx$ conductivity looks more flat than $\sqrt{\Omega}$ in behaviour.  Another feature is that the Drude conductivity does not depend on $\alpha$ at all. It is unchanged. For the Fig.~\ref{fig12} plots, the dashed curves are from numerical evaluation of
 Eqs.~(\ref{eq:kylexx}) and (\ref{eq:kyleyy}) with $\Gamma=0.007$ in the broadened spectral functions $A^{0,\pm}(\omega)$ and $A^{0,\pm}(\omega+\Omega)$.  
The solid curves are those using our analytical formulas Eqs.~(\ref{eq:at3intraxx})-(\ref{eq:at3interyy}). To implement the intraband piece of Eqs.(\ref{eq:at3intraxx}) and (\ref{eq:at3intrayy}), we used the Lorentzian representation of the delta function and used a broadening of $2\Gamma$ as the Lorentzian resulting from the convolution of two Lorentzians in the conductivity Kubo formula, each of width $\Gamma$, will have a width of $2\Gamma$. The agreement between the analytical formulas and the numerics is excellent and illustrates the effect of the impurity smearing on the absorption edges. It is clear that observing two absorption edges, one at a frequency that is twice the other, would be a possible indication of the presence of a flat band in a system with non-integer $\alpha$. Moreover, the different frequency dependence from polarization in the $x$ versus the $y$ direction, as discussed here, would be  a signature of a semi-Dirac system. Flat bands or nearly flat bands are currently of high interest due to the potential of exotic physics arising from the high degeneracy of such bands.

\section{Summary and Conclusions}

The optical properties of a semi-Dirac material in the clean limit differ in many aspects from those of graphene but in others are very similar. The graphene conductivity is isotropic while in the semi-Dirac case, the longitudinal conductivity along the $x$ and $y$  direction can be very different from each other. In particular, the material parameters $m$ and $v$ which define the dependence on momentum of the dispersion curves, {\it i.e.} quadratic (nonrelativistic) $\hbar^2k^2_x/2m$ in the $x$-direction and linear (relativistic) $\hbar vk_y$ in the $y$-direction, do not drop out of $\sigma^{xx}$ and $\sigma^{yy}$ but enter inversely to each other so that they cancel in the square root of the product $\sqrt{\sigma^{xx}\sigma^{yy}}\equiv\sigma$. The dependence on photon energy $\Omega$ is also quite different in the two directions,  $\sqrt{\Omega}$ in $\sigma^{xx}$ 
as opposed to $1/\sqrt{\Omega}$ in $\sigma^{yy}$ for interband transitions. On the other hand $\sigma$ is independent of $\Omega$, $m$ and $v$, with the magnitude of the constant background slightly different from that in graphene. A similar situation is obtained for the intraband (Drude) part of the conductivity which goes as the square root of the chemical potential for $yy$ and as $\mu^{3/2}$ for $xx$, with the square root of the product being linear in $\mu$ and universal like in graphene but with a modified magnitude in units of $e^2/h$. However, for
$\sigma^{xx}$ no sumrule applies to the optical spectral weight transfer from the interband background to the Drude peak as doping is increased, while it still applies to $\sigma^{yy}$. In semi-Dirac, the optical spectral weight in the Drude is larger than the amount lost in the interband background which is smaller by a factor of 2/3, independent of the value of the chemical potential when the quantity 
$\sqrt{\sigma^{xx}\sigma^{yy}}$ is considered.

For a general photon energy, with the introduction of impurity scattering characterized by a residual rate of $\Gamma$, the Dirac delta function of the clean limit Drude is broadened and the hole (due to Pauli blocking) in the interband optical conductivity extending  from $\Omega=0$ to $2\mu$ (in the clean limit) is partially filled in but remains clearly identifiable provided $\Gamma<<\mu$. These effects are superimposed on the simple $\sqrt{\Omega}$ and $1/\sqrt{\Omega}$ behaviors of the clean limit which remain in the region of $\Omega>>\mu$. In the dc limit $(\Omega\to 0)$, $\sigma^{yy}$ is found to be finite and proportional to $1/\sqrt{\Gamma}$ while $\sigma^{xx}$ goes like $\sqrt{\Gamma}$. These replace the divergent response of the clean limit $\sigma^{yy}$ and zero dc value of  $\sigma^{xx}$. At finite $\Omega$, the square root of the product $\sigma^{xx}(\Omega)\sigma^{yy}(\Omega)$ is found to be very close in qualitative behavior to previous results for graphene although there are some quantitative differences. In particular, the  dc limit of this quantity goes as $0.398 e^2/h$, while in graphene it is $e^2/(\pi h)$, when the degeneracy factor is $N_f=1$ for comparison.

For the clean limit, we provide results when a on-diagonal gap is introduced in the Hamiltonian. The optical gap edge at $\Omega=2\Delta$ is found to vary as $[\Omega^2-(2\Delta)^2]^{1/4}$ for $\sigma^{xx}_{\rm inter}(\Omega)$ and as the inverse for
$\sigma^{yy}_{\rm inter}(\Omega)$. These dependences cancel in $\sigma_{\rm inter}=\sqrt{\sigma^{xx}_{\rm inter}(\Omega)\sigma^{yy}_{\rm inter}(\Omega)}$. For the intraband contribution, $\sigma^{xx}_{\rm intra}(\Omega)$ is proportional to 
$(\mu^2-\Delta^2)^{5/4}/\mu$ and $\sigma^{yy}_{\rm intra}(\Omega)$ to $(\mu^2-\Delta^2)^{3/4}/\mu$, so the square root of their product goes like 
$(\mu^2-\Delta^2)/\mu$  and this reduces to linear in $\mu$ when the gap $\Delta=0$.
  
 We have considered a generalization of the semi-Dirac case to include arbitrary positive integer powers of $k_x$ and $k_y$, $n$ and $s$, respectively. In this case, the electronic density of states $N(\omega)$ 
is found to vary as
$N(\omega)\sim \omega^{{1\over n}+{1\over s}-1}$  and 
$\sigma^{xx}_{\rm inter}(\Omega)\sim\Omega^{{1\over s}-{1\over n}}$ and  $\sigma^{yy}_{\rm inter}(\Omega)\sim\Omega^{{1\over n}-{1\over s}}$, the inverse dependence. These dependencies are constant for any value of $n$ if $n=s$ in which instance
 $N(\omega)\sim \omega^{{2\over n}-1}$  and only for $n=2$ is this quantity constant. It is linear for $n=1$ (graphene). For $n=2$ and $s=1$ (semi-Dirac), it is $\sqrt{\Omega}$. These special cases are known results. $\sigma_{\rm inter}=
\sqrt{\sigma^{xx}_{\rm inter}  \sigma^{yy}_{\rm inter}  }$ is always constant for any value of $n$ and $s$, but its magnitude does depend on $n$ and $s$.

A generalization to include an additional flat band using the $\alpha$-${\cal T}_3$ semi-Dirac model has also been considered. It is still possible to obtain analytic results for the intraband and interband contributions to the real part of the conductivity in the clean limit. These are tested against numerical results that include
 a small residual scattering rate.

In conclusion, by examining various models demonstrating semi-Dirac behavior and also considering a generalization of the basic Hamiltonian to higher power-laws, we have provided analytical formulas for the frequency-dependent optical conductivity and dc conductivity which may be used to identify and confirm semi-Dirac physics in new materials. As the optical conductivity technique has played an important role in the study of new materials and in particular Dirac materials, we anticipate that these results will assist in advancing the field.

\appendix*
\section{Angular integrals}

Throughout our calculations a number of angular integrals are encountered and we summarize them here to assist those who may wish to reproduce our results or do further work. Following the work of Pi\'echon et al.\cite{Piechon:2015} on the semi-Dirac model, we transform variables from $(k_x,k_y)\to (\epsilon,\phi)$ via $\hbar^2k_x^2/2m = \epsilon\cos\phi$ and $\hbar vk_y = \epsilon\sin\phi$,
where $\phi$ is restricted to the interval $(0,\pi/2)$.  This transformation on the sum over ${\bm k}$ along with the transformation of the integrands provides a series of integrals which can be written in terms of the Gauss constant:

\begin{equation}
G={2\over \pi}\int_0^1 {dx\over\sqrt{1-x^4}}\approx 0.8346
\end{equation}

Integrals which appear in the evaluation of the $xx$ quantities are:
\begin{eqnarray}\displaystyle
 &\displaystyle\int^{\pi/2}_0\sqrt{\cos\phi}\cos^2\phi d\phi = {3\over 5G},\\
&\displaystyle\int^{\pi/2}_0\sqrt{\cos\phi}\sin^2\phi d\phi ={2\over 5G},
\label{eq:Cxxs}
\end{eqnarray}
which add to give
\begin{equation}
 \displaystyle\int^{\pi/2}_0\sqrt{\cos\phi} d\phi ={1 \over G}.
\end{equation}
Those which appear in $yy$ quantities are:
\begin{eqnarray}\displaystyle
 &\displaystyle\int^{\pi/2}_0{\cos^2\phi\over\sqrt{\cos\phi}}d\phi = {\pi G\over 3},\\
&\displaystyle\int^{\pi/2}_0{\sin^2\phi\over\sqrt{\cos\phi}} d\phi ={2\pi G\over 3},
\label{eq:Cyys}
\end{eqnarray}
which combine to
\begin{equation}
 \displaystyle\int^{\pi/2}_0{1\over\sqrt{\cos\phi}} d\phi ={\pi G}.
\end{equation}

\begin{acknowledgments}

This work has been supported by the Natural Sciences and Engineering Council of Canada (NSERC) and by the
Canadian Institute for Advanced Research (CIFAR).

\end{acknowledgments}


%

\end{document}